\documentclass[aps,amsmath,twocolumn,nofootinbib]{revtex4-1}
\usepackage{graphicx}
\usepackage{relsize}
\usepackage{enumitem}
\usepackage[table,usenames,dvipsnames]{xcolor}
\usepackage{ifthen}
\definecolor{linkcolor}{rgb}{0.6,0,0}
\definecolor{citecolor}{rgb}{0,0,0.75}
\definecolor{urlcolor}{rgb}{0.12,0.46,0.7}
\usepackage[breaklinks, colorlinks, urlcolor=urlcolor, linkcolor=linkcolor,citecolor=citecolor,pdfencoding=auto]{hyperref}
\hypersetup{linktocpage}
\setlist{nolistsep,leftmargin=*} 

\begin{document}


\title{Science Spoofs, Physics Pranks and Astronomical Antics}

\author{Douglas Scott} \email{docslugtoast@phas.ubc.ca}
\affiliation{Dept.\ of Physics \& Astronomy,
 University of British Columbia, Vancouver, Canada}

\date{1st April 2021}

\begin{abstract}
Some scientists take themselves and their work very seriously.  However, there
are plenty of cases of humour being combined with science.  Here I review
some examples from the broad fields of physics and astronomy, particularly
focusing on practical jokes and paper parodies.  This is a mostly serious
overview of a non-serious subject, but I'd like to claim that there is in fact
some connection between humour and creativity in the physical sciences.
\end{abstract}

\maketitle

\tableofcontents

\section{Scope}
The physical sciences are usually taken to be very earnest pursuits by
those who work in
them.  However, most professional physicists and astronomers would also
happily agree with the t-shirt mantra that says ``physics is
phun'' \cite{tshirt}.
I suspect that anyone who finds it no fun to understand the nature
of physical reality will never end up pursuing it as a
career.  All professionals surely have seen some humour in the subject
matter of their job.  Hence, in addition to the
weighty aspects of the physical sciences, there is a lighter side,
which is the focus of this paper.

This topic has been written about before.  There are many examples
of the use of humour in physics and astronomy, making it impossible to
give a comprehensive review.  Instead I focus on key examples of
deliberate jokes and attempts to use wit in the physical sciences.
There will be some bias towards areas closer to my own research in cosmology
and astrophysics, but I've tried to include examples across the broad
disciplines of physics and astronomy.

To describe the scope of this article, it is useful to
explain what I am {\it not\/} going to discuss.
Deliberate deceptions will not be my focus, e.g.\ proposals for perpetual
motion machines or anti-gravity devices.
I will not cover sensational phenomena that resulted from delusions or
systematic errors rather than humour.  Examples here include claims for:
the existence of the planet Vulcan, 1859; Martian ``canals'', 1877;
``N-rays'', 1903; rotation of spiral nebulae, 1916;
premature verification of gravitational redshift, 1925; E.S.P., 1934;
Dogon astronomy knowledge, 1948; ``Worlds in Collision'', 1950;
``polywater'', 1961; cosmic-ray-created superpowers, also 1961;
detection of gravitational waves, 1969;
anomalous behaviour of a pendulum during a solar eclipse, 1970; 
the ``Oops-Leon'' particle, 1976;
the ``17-keV neutrino'', 1985;
a fifth force, 1986;
``water memory'', 1988; ``cold fusion'', 1989; faster-than-light
neutrinos, 2011; ``perytons'' (that turned out to be microwave ovens), also
2011; and primordial cosmic-microwave-background ``$B$ modes'', 2014.

Additionally I will steer clear of most
claims that involved scientific misconduct.  An example is
the ``Sch{\"o}n scandal'' in condensed-matter physics, which arose from
claims by Jan Hendrik Sch{\"o}n that he could make
single-molecule semiconductors and nano-scale circuits using
organic materials.  Another case is the alleged
fabrication of data leading to the claimed discovery of element 118 in 1999.
There are other examples with much controversy over whether
or not misconduct occurred, e.g.\ the theoretical cosmology papers of the
Bogdanov brothers and several plagiarism scandals in other science fields.
I will not mention any of these examples, instead sticking with deliberate
pranks and related antics.

\section{Why be funny?}
There's certainly nothing to stop anyone mixing
physics (or astronomy) with comedy, although
``physical comedy'' means something entirely different of course!  And
we're all aware of a TV sitcom starring a theoretical physicist, an
experimental physicist and an astrophysicist (plus an engineer, and some
life scientists too).  But
is there any actual connection between physics and humour?

In 1957 the English physicist R.V.
Jones published ``The theory of practical joking -- its relevance to physics''
\cite{Jones}.  The article starts ``At first sight there may seem little
relation between physics and practical joking. Indeed, I might never have
observed their connection but for an incidental study of the life of
James Clerk Maxwell \dots''  Jones goes on to describe how the use of analogy
and incongruity are common to both humour and physics.  He points out that,
as well as Maxwell, there have been other well-known jokesters in physics,
including George Gamow (who will reappear later).
Jones himself carried out so-called ``phone pranks''
and credits German scientist Carl Bosch
as the originator of the trick whereby someone is convinced that
they can be seen through their telephone -- a prank he says Bosch pulled on a
journalist staying across the street from him in about 1933 \cite{Bosch}.
It is worth pointing out that Jones worked on ways to fool enemy radar during
World War~II and is sometimes called the ``father of scientific intelligence''.

The connection between physics and humour is developed even further in
a recent paper, ``Toward a Quantum Theory of Humor'' \cite{quantum}.
The authors dissect the joke ``Time flies like an arrow,
but fruit flies like a banana'' and attempt to model it as sets of
quantum states, where the wavefunction of ambiguous framings of the joke 
is collapsed by the measurement process into ``funniness'' states.
It should be stressed that the paper itself is not a joke!

In physics education research, the recent paper ``The Role of Humor in
Learning Physics: a Study of Undergraduate Students'' \cite{PER} shows that
humour might contribute to a good work atmosphere, and hence improve
learning outcomes.  The study also suggests that, through humour,
students can find pathways to engage in discourse within physics.  Similarly,
a study carried out in Portugal found that audiences appreciated stand-up
comedy routines developed by scientists \cite{standup}.  On the
other hand, a study in 1998 \cite{planetarium} showed that viewers
of a humour-laden planetarium show retained less information
than those who saw a non-humorous show!

Nevertheless, I suggest that, used carefully, humour in the classroom
is a good thing.  There's certainly nothing worse than instructors who
don't appear to be enjoying themselves!  In research too, it's important
to see the funny side.  I often think that some of my colleagues
take themselves far too seriously.  They seem to have lost sight of the
fact that, in most areas of physics and astronomy, the only real reasons for
pursuing research are interest and enjoyment (including public and
student engagement of course).  If attacked
for working on cosmology because it's inherently useless, I like to respond
that while it may be useless, at least it's also harmless!

Much has been written about the nature of humour; probably the only
ingredient agreed on is that things tend to be funny because of some
dissonance or incongruity.  Often humour, particularly parody,
plays an important role in showing the absurdity in a situation, or
simply bringing the over-serious back down to Earth \cite{Earth}.

\section{Publications devoted to humour}
Several existing books include source material
on humour in physics and astronomy.  ``A Random Walk in Science'' (1973)
\cite{random1} and its sequel ``More Random Walks in Science'' (1982)
\cite{random2} are collections of light-hearted
contributions, mostly about physics, but also including items from other
sciences.  Although some pieces now seem a bit dated, on the whole these books
are excellent and refreshing collections that I recommend highly.
An earlier compilation from the Soviet Union, called
``Physicists Continue to Laugh'', was published in 1968 \cite{MIR},
but there has never been a full English translation.

In 1969 the UNESCO-supported journal ``Impact of Science on Society'' had a
special issue on ``The science of humour, the humour of science''
\cite{impact}.
Another anthology with some content related to physics and astronomy
is ``Laughing Space'', edited in 1982 by Isaac Asimov and Janet Jeppson
\cite{Laughing}.  ``Absolute Zero'' is a 1992 compilation of jokes and
anecdotes related to science and scientists, collected by Betsy Devine and
Joel E. Cohen \cite{AZGravity}.  ``Science Askew'' by Donald Simanek and
John Holden, published in 2001, contains satirical articles, as well as
jokes, puns, stories and quotes, many related to physics \cite{Askew}.
``Academia Obscura'' by Glen Wright
includes some relevant material, but mainly focuses on the life sciences
\cite{Obscura}.  Jan Witkowski wrote a series of magazine articles about humour
in science, also concentrating on biology \cite{Witkowski}.
A new book contains a couple of chapters that cover some of
the same ground as this review \cite{Booknote}: ``Fake Physics:
Spoofs, Hoaxes and Fictitious Science'' by Andrew May \cite{May}.

The Cavendish Laboratory in Cambridge, had a tradition of humorous
songs sung by students after the annual dinner.  Words were specially
written to tunes from Gilbert \& Sullivan and other popular music sources.
The programmes were printed privately in 1904, 1906, 1907 and 1911,
then published more
formally in 1920 and 1926 as ``Postprandial Proceedings of the Cavendish
Society'' \cite{Prandial}.  Many of the songs were written by ``A.A.R.'',
physicist Alfred A. Robb, who worked under J.J.\ Thomson at the Cavendish
and was later known for his books on special relativity.  As an illustration,
here's part of what Robb wrote about Thomson:
\par
{\sl What's in an atom}
\par\quad
{\sl The innermost substratum?}
\par
{\sl That's the problem he is working at today.}
\par
{\sl He lately did discover}
\par\quad
{\sl How to shoot them down like plover,}
\par
{\sl And the poor little things can't get away.}
\par\noindent
In another example Gilbert Stead (later known as a pioneer of radiology)
put words about Planck's law to the tune of ``Men of Harlech'', with the
following first verse:
\par
{\sl All black body radiations,}
\par
{\sl All the spectrum variations,}
\par
{\sl All atomic oscillations,}
\par
{\sl Vary as $h\nu$.}
\smallskip

The Journal of Jocular Physics was a spoof journal produced at the Institute
for Theoretical Physics in Copenhagen as a tribute to Niels Bohr,
for his 50th (1935), 60th (1945) and 70th (1955) birthdays
\cite{CERNArchive}.  Contributors included
L{\'e}on Rosenfeld, Victor Weisskopf, George Gamow, Oskar Klein
and Hendrik Casimir.  This same institute (now named after Bohr) was also
known for its informal conferences that ended with comedic skits.

The Journal of Irreproducible Results (JIR) was started in Israel in
1955 by virologist Alexander Kohn and physicist Harry J. Lipkin.
It is still published, despite several changes in who runs it \cite{JIR}.
In 1994 JIR Editor Marc Abrahams left to found the
rival journal Annals of Improbable Research (AIR) \cite{AIR}, which is also
connected with the Ig Nobel Prizes.  Both the JIR and AIR publications are
devoted to
scientific humour and there have been many examples related to physics
and astronomy (although AIR leans a little more toward life sciences).

Ig Nobel Prizes are given annually to reward studies ``that first make people
laugh, and then make them think''.  It is regularly
given for Physics and more occasionally for Astronomy.
Winners have included scientists
who tried to answer the following questions:
``why is it so easy to slip on a banana peel?'';
``could humans walk on water on the Moon?'';
``how do knots form in jostled string?'';
``what's the longest continuously running laboratory experiment?'';
and ``can I magnetically levitate a frog?''  A more complete list is
given in Appendix~A.

The Worm Runner's Digest was started by a biologist in 1959 and ran for
20 years \cite{Worm}.
It published both satirical and serious articles, with the jokey
ones printed upside down once it became clear that some people found it hard
to tell the difference!  Null Hypothesis: The Journal of Unlikely Science is
an on-line website based in the U.K., which started as a magazine in 2004.
Founded by three biology graduate students, it contains content from a wide
range of topics, including the physical sciences.  There are also many
science-related stories on internet-based humour sites, such as ``The Onion''
\cite{Onion}.

In addition to these specifically humour-based publications,
several other journals and newsletters occasionally include
non-serious contributions.  Applied Optics had a section called ``Of Optics
and Opticists'' that occasionally contained humorous pieces, and there were
similar contributions in
``N.P.L.\ News'' from the U.K.'s National Physical Laboratory.
Also in the U.K., ``The Observatory'' is a bimonthly
review of astronomy, which has been published since 1877; in addition to
serious papers and reports, it includes a ``Here and There'' section,
pointing out misprints and ridiculous statements of astronomical interest.
On three specific occasions, there has been a 
``Special Pull-out and Throw-away Supplement'' added; these were to commemorate
the 1000th issue (in 1974), the 100th year (in 1977) and the year 2000 (in
2000, naturally) \cite{Observatory}.
The supplements were ``pink pages'', perhaps to mimic the
old letters section of Monthly Notices of the Royal Astronomical Society;
they contained more frivolous content, including reports of fake
meetings, poetry and spoofs of papers, e.g.\ ``On the properties of cuboid
star clusters'' \cite{Cuboid}, ``On the possible existence of the lost
constellation of `Cuculus' (the Cuckoo)'' \cite{Cuculus}
and ``Astrophysics in 2049'' \cite{Trimble}.

\section{Early examples of science jokesters}
The tradition of combining the study of science with humour goes back at
least to the ancient Greeks.  Theophrastus, Aristotle's successor at
the Athens Lyceum, wrote about the properties of the natural world
(with fragments of his ``History of Physics'' surviving), as well as writing
humorously about personality traits (in the work ``Characters'')
\cite{Theophastus}.

More than a thousand years later,
the scholar Michael Scot lived from 1175 until about 1232 \cite{MScot}.
He studied in several of the
great centres of learning in Europe and was court astronomer (or astrologer)
to Frederick~II of Sicily.  He translated Aristotle into Latin,
including the book on astronomy and related topics, ``De Caelo''.
Fibonacci dedicated one of his works to Scot, and Scot
may have been the first person to describe the phenomenon of multiple rainbows.
In ``Super auctorem spherae'', Scot gives a dialogue about astronomy
between a wise man and a simpleton referred to as ``Sir Lupus Fiat'', which is
an anagram of ``Aprilis Fatuus'' (Latin for April Fool).  Albeit indirect,
this may be one of the first mentions of the connection between practical
jokes and the month of April; interestingly, it occurs in a treatise
on astronomy.

Galileo Galilei has a reputation as the world's first serious physicist and
astronomer, but he also had his waggish side.
One thinks of his ``Dialogue Concerning the Two Chief World Systems'' (actually
a trilogue) \cite{Dialogue}, where the conservative-thinking ``Simplicius''
(named after a 6th-century commentator of Aristotle) is
represented as a slow-witted fool.  When he was a young man,
Galileo wrote some sonnets and also a diatribe against the poet Tasso.
Additionally he wrote two satirical poems: ``In Abuse of Gowns'', poking fun at
the rules that professors in Pisa had to wear academic robes; and
``It's All Relative'', describing how the famous tower was in fact straight,
but everything else was leaning.

The writings of the great satirist Jonathan Swift contain several passages
related to the sciences.  In Part~III Chapter~5 of ``Gulliver's Travels''
he describes ``The Academy'' on Lagado, which is essentially a mocking attack
on the apparent uselessness of some academic studies \cite{Swift}.
He specifically describes pointless experiments, for example to extract
sunbeams from cucumbers.  He also wrote that the scientists of Laputa had
discovered ``two lesser stars, or satellites, which revolve about Mars'',
coincidentally fitting the discovery of Phobos and Deimos 150 years later.
Swift was additionally the originator of a prank at the expense of a
contemporary astrologer, John Partridge, who published a series of predictions
in 1708, including the deaths of several prominent people.  A pamphlet
quickly appeared, written by Isaac Bickerstaff, containing the prediction that
Partridge himself would die on 29th March.  This was followed by another
pamphlet on 30th March claiming that Partridge had in fact died as
predicted -- which would have been read by many people on 1st April of course.
Isaac Bickerstaff was a pseudonym for none other than Jonathan Swift
\cite{Bickerstaff}.

Newton famously said ``If I have seen further it is by standing on the
shoulders of Giants''; the quote is so well known that it was engraved on
the edge of the British \pounds2 coin.  The sentence occurs in
a letter that Newton sent to his rival Robert Hooke.
Some recent historians of science \cite{Chapman}
have suggested that this was a deliberate
dig by Newton at the expense of Hooke, who was described as being
small of stature and with pointed features.  Certainly we know that
Newton grew to regard Hooke as an enemy.  Indeed in several letters
he referred to Hooke's most
famous discovery as ``Hook's Law'' rather than ``Hooke's Law'' \cite{Hook},
mocking the facial features of his fellow physicist.  It could therefore be
said that he picked on the nose of his rival.

Benjamin Franklin used humour to write about electricity in newspaper articles
and regularly played pranks on visitors to his home by giving them mild
electric shocks \cite{Franklin}.
Michael Faraday used humour in his popular lectures, including the series
on the ``Chemical History of the Candle'', which started the tradition of the
Royal Institution Christmas Lectures.  {\it And\/} he invented party balloons!

An example of early humour that is well known to chemists was a pair of spoof
papers written by Justus von Liebig and Friedrich W{\"o}hler in 1839 and 1840.
These gentlemen used their positions as editors to publish ``Spirit and
Ferment: The Mystery Dispelled'' \cite{Chemistry1},
ridiculing the idea that fermentation was a biological (not chemical) process
and ``On the Substitution Law and the Theory of Types'' \cite{Chemistry2},
mocking the claim that a substance would retain its properties even when some
of its atoms were replaced by those of another element.  The first was
apparently an anonymous letter to the editor, but the second was written by
S.C.H.~Windler \cite{Chemistry3}.

In 1886 there was a whole spoof edition of the Berichte der Deutschen
Chemischen Gesellschaft called ``Berichte der Durstigen Chemischen
Gesellschaft'' (Journal of the Thirsty Chemical Society).
This contained an article with a figure showing interlocking monkeys
in a circle as the structure of benzene \cite{monkeys}.
Whether this was a parody of the famous ``snake
swallowing its tail'' dream image, or if instead this spoof contributed
to later retellings of the story, remains debated by science historians.

James Clerk Maxwell, the great Scottish physicist of the 19th century,
was known as a prize prankster.  It is said that in 1871 he
arranged for his inaugural professorial lecture at Cambridge to be
advertised only to undergraduates, while the Fellows and Dons of the
university came instead to the first lecture in his undergraduate course,
where he explained things
like the difference between Fahrenheit and Centigrade.  Maxwell also
wrote many amusing poems that mixed science with creative
writing, e.g.\ in ``Report on Tait's Lecture on Force'' \cite{MaxwellPoem}:
\par
{\sl Force, then, is Force, but mark you! not a thing,}
\par\quad
{\sl Only a Vector;}
\par
{\sl Thy barb{\`e}d arrows now have lost their sting,}
\par\quad
{\sl Impotent spectre!}
\par
{\sl Thy reign, O Force! is over. Now no more}
\par\quad
{\sl Heed we thine action;}
\par
{\sl Repulsion leaves us where we were before,}
\par\quad
{\sl So does attraction.}
\par\noindent
Thus Maxwell used a witty verse to describe how the physical picture of
interactions had shifted from forces to fields.

Paul Dirac, one of the most prominent theoretical physicists
of the 20th century, was famous for being taciturn and {\it not\/} known for
his sense of humour.  The following lines of verse are attributed to him:
\par
{\sl Age is, of course, a fever chill}
\par
{\sl That every physicist must fear.}
\par
{\sl He's better dead than living still}
\par
{\sl When once he's past his thirtieth year.}
\par\noindent
And yet he is also quoted as saying ``In science one tries to tell people,
in such a way as to be understood by everyone, something that no one ever knew
before. But in the case of poetry, it's the exact opposite!''  However,
perhaps Dirac's sense of humour was just more mysterious than most?  In 1939
he published a 2.5-page paper describing a new notation for representing
quantum states \cite{braket},
which he ended with the sentence ``As names for the new symbols
$\,\left\langle\right.\,$ and $\,\left.\right\rangle\,$ to be used in speech,
I suggest the words {\it bra\/} and {\it ket\/} respectively'' \cite{notket}.

George Gamow, the prominent Ukrainian-American theoretical physicist of the
mid-20th century, was known for his humour as well as his contributions to
cosmology and other fields.  This trait can be seen in his popular
``Mr.\ Tompkins'' books,
as well as in his naming the neutrino ``Urca process'' after a casino,
and in the many pranks he carried out.  He once submitted a paper to Nature
claiming that an explanation for cows chewing clockwise versus anticlockwise
in different hemispheres lay with the Coriolis force and he tried (but
sadly failed) to get a paper accepted with Mr.\ Tompkins as a co-author
\cite{Tompkins}.
Gamow also wrote verse, for example this one in 1964 about the newly
discovered ``quasi-stellar objects'':
\par
{\sl Twinkle, twinkle, quasi-star,}
\par
{\sl Biggest puzzle from afar.}
\par
{\sl How unlike the other ones,}
\par
{\sl Brighter than a billion suns.}
\par
{\sl Twinkle, twinkle, quasi-star,}
\par
{\sl How I wonder what you are.}
\vspace{0.1cm}

Anthony P. French, known for his four classic physics textbooks, also
wrote humorous poems about the history of his discipline.  He specifically
created limericks and ``double dactyls'' \cite{DD}, including this one about
quantum mechanics:
\par
{\sl Higgledy, piggledy}
\par\quad
{\sl Erwin `H' Schr{\"o}dinger,}
\par\quad
{\sl Said `Here's a recipe --}
\par\quad
{\sl Don't ask me why}
\par
{\sl Structures of atoms are}
\par\quad
{\sl Fully described by a}
\par\quad
{\sl Quantum-mechanical}
\par\quad
{\sl Function called $\Psi$'.}

\section{Name mix-ups}
Surnames can have different spellings, especially in previous centuries
(as we already saw with Hooke).
An early example of name confusion in physics comes when one studies
electromagnetism and related areas and realises that
some effects are attributed to Lorentz and others to Lorenz.  It is
a surprise to many when they realise that there appear to be two different
scientists here.  But the truth is even stranger.  In fact
Hendrik Ludvig Lorentz used alternative spellings at different periods
and in different journals \cite{Lor}.  When discussing a way to fix the
electromagnetic vector potential he adopted the form ``Lorenz'' for the gauge
condition.  He used both names when he derived the relationship between the
refractive index and the density of a medium.  But later he is ``Lorentz''
for his publications on the Lorentz force and the Lorentz transformation.

A similar prank was pulled by David Marsh in his 1st April 2019 paper
``The Marshland Conjecture'' \cite{Marsh}.
The main claim is that there are in fact
two separate David Marshes, ``David M.C. Marsh'' and ``J.E. David Marsh'',
each writing papers in overlapping areas.  Most
readers seemed convinced by the detailed description in the paper of these two
parallel careers (while in reality they are both the same person),
making this a very successful April Fool!

\section{Conflating degrees with degrees}
Sometimes a joke paper can be written to explicitly mock some
previous publication.  There's probably a unique instance where a paper
of this sort was actually published in a reputable journal.  This example
is ``On the Quantum Theory of the Absolute Zero'', written by Beck, Bethe
\& Riezler in December 1930 and published in Die Naturwissenschaften in
January 1931 \cite{degrees}.

The motivation was a calculation by Eddington yielding an explanation for the
value of the inverse fine-structure constant
$1/\alpha\simeq137$.  This was essentially a piece of numerology, counting the
elements in a symmetric $16\times16$ matrix, plus 1 for the orbital motion of
an electron \cite{Eddington}.
Beck, Bethe \& Riezler were all postdocs in Cambridge when
they heard Eddington give a lecture on this topic and were inspired to create
their lampoon.  In their own contribution they discuss the number of degrees of
freedom in a crystal and how to reach absolute zero temperature.  The short
Beck et al.\ paper cleverly transitions from ``degrees of freedom'' to
``degrees of temperature'' so that a casual reader might not notice the switch.
Then they show that the
absolute zero level of $-273^\circ$ comes from $-(2/\alpha-1)$.

The paper was published in a serious German journal \cite{German},
but the editor Hans Spemann was not amused when he
found out that it was a joke.  So a retraction was published a couple of
months later, with the authors said to ``express regret that the formulation
they gave to the idea was suited to misunderstanding''.
One doubts whether in fact they regretted it at all!

\section{April Fools}
For centuries the first day of April has been known in many western countries
as ``April Fool's Day'' (or its equivalent).  Its origins are unknown, but
there have been suggestions that the practice goes back to the Roman festival
of Hilaria, which was at about the same time of year.
It is a day set aside for playing tricks
on one's friends, the tricks normally being of the non-malevolent kind,
such as sending someone on a fool's errand.
One of the earliest recorded April Fool's Day pranks was in 1698 when it
was announced that people could come to the Tower of London to see the
lions being washed, while no such ceremony actually took place \cite{lions}.

\subsection{April Fools in the media}
There are many examples of pranks propagated through
newspapers, radio, TV and the internet on 1st April.  Famous examples
include spaghetti growing on trees, flying penguins and a burger for
left-handed people.  Below are some examples more specifically
related to physics and astronomy.
\begin{itemize}
\item
Energy was harnessed from the atmosphere according to a 1923 article in
the newspaper Deutsche Allgemeine Zeitung; the story was picked up seriously by
the New York Times, the LA Times and others.
\item
There was an atomic blueprint scare in London in 1952.
\item
A discussion of ``contra-polar energy'' appeared in Popular Electronics in
1955; this could remove light from an affected area, for example.
\item
In 1959 it was announced that the twin satellites of Mars were artificial
\cite{MarsMoons}.
\item
A Swiss Moon-landing hoax of 1967 caused people to flock up mountains to get
a better view for themselves.
\item
Metric time was introduced in Australia in 1975.
\item
In 1984 it was announced that light is caused by an absence of ``darkons''.
\item
So-called ``bigon'' particles, which are the size of bowling balls,
were discovered in 1996.
\item
``Guinness Mean Time'' replaced Greenwich Mean Time at the Royal Observatory
in 1998.
\item
A teleportation machine was invented in 2013.
\item
In 2015 CERN confirmed a fifth type of fundamental interaction, called
``The Force''.
\item
Also in 2015, Scientific American announced they would be abandoning
April Fool's Day \cite{SciAm}.
\item
A company produced a Flat Earth Globe in 2019.
\item
Astronomers declared in 2021 that we've learned everything we need to
know about the Universe.
\item
The use of a quantum computer to run a Zoom session in 2021 resulted in
participants being in superpositions of breakout rooms.
\end{itemize}

\vspace{0.1cm}
The NASA ``Astronomy Picture of the Day'' (APOD)
website has regularly presented joke
images on 1st April.  Perhaps the most successful of these was in 2005,
when on 31st March the site contained the teaser ``Water on Mars!''\@ for its
presentation to follow on the next day; the results of the wait disappointed
many \cite{WaterMars}.  Further examples are listed in Appendix~B.

\subsection{Litre April Fool}
Many science-related pranks have been carried out on this particular day,
or at least spoof papers have been submitted with this date.  An outstanding
example regarding units appeared in the April 1978 issue of ``CHEM 13 News'',
a newsletter for high-school teachers.  Ken Woolner, a physicist from
the University of Waterloo, suggested that the litre (whose symbol is often
``L'', since ``l'' is easily confused with the number ``1'') is named after
Claude {\'E}mile Jean-Baptiste Litre (1716--78) \cite{Litre}.  Apparently
Litre was the son of a wine-bottle manufacturer and later worked as a
creator of precision instruments, proposing a new unit of volume that
was later adopted by the Syst{\`e}me international.  The idea for the
biography originally
came from Woolner's colleague Reg Friesen, during a blizzard when the pair
were stuck in a hotel room in Ottawa.  Woolner wrote out a detailed life
history for Litre, and to add colour to his article he wove in elements of
French history and scientific luminaries of the time.  For
authenticity he left some gaps since ``the details of Litre's life are very
hard to establish , and most of this account was inferred
from the general literature of the period''.
Some scientists joined in with the joke, filling in some of the missing
pieces of the biography, including that Litre had a daughter called Millie.
Later, several published descriptions of the S.I. units accepted the
Litre story as fact, and eventually Woolner had to come clean about his
joke \cite{LWiki}.
As a last comment, this story is likely known more widely than Ken Woolner's
more serious contributions to science!

Woolner and Friesen may well have been inspired by an earlier story from
the newsletter of the U.K. N.P.L.
\cite{Moire}.  This concerns Jean-Baptiste Moir{\'e}, who gave his name to the
Moir{\'e} pattern.  The pseudonymous author ``Simplicius'' \cite{Simplicius}
gives a brief biography of
Moir{\'e}, connecting him to Champollion, the Pre-Raphaelites and Japanese
silk-weavers, as well as to Prof.\ Eddy (discoverer of currents) and George
Canap{\'e} (the renowned chef).

This Moir{\'e} story may itself have been inspired by an earlier report
from the 1962 JIR about how Juan Hernandez Torsi{\'o}n Herrera gave his name
to the Torsion balance, apparently when watching his cousin being tortured on
the rack \cite{Torsion}.  The authors were ``Col.\ Douglas Lindsay and
Capt.\ James Ketchum'' \cite{Ketchum}.  And no doubt there are even earlier
examples of similar spoof science histories.

The Torsion and Moir{\'e} stories are recounted along with several others in
the 1996 article ``Some Famous Names in Physics'' by Australian physicist
Tony Klein \cite{Klein}.  The stimulus
was a talk by Swedish physicist (and Nobel Committee member) G{\"o}sta Ekspong
(who changed his name from Carlson, {\it after\/}
publishing what was perhaps his most
famous discovery).  Klein mentions many additional physicists who are
eponyms, including Matthew Fringe, Wolfgang Bremsstrahlung,
Emilio Carburetto, Hercule Parallax, Claude Neon, Katherine Scanning, Jesus
Klystron, Spiros Solenoides and the Coupling brothers, J.J. and L.S.

\subsection{April Fool's papers}
Returning to the topic of 1st April, a tradition has grown of submitting
joke papers to the preprint arXiv on this day.
These are mostly not ``April Fools'' in the true sense, since typically
they are so outlandish in their claims that they fool no one.  But as examples
of humorous parodies of papers, they are submitted in a similar spirit to the
old celebrations of this day.
A list of known arXiv joke papers submitted on 1st April (or around then,
since sometimes it is hard to judge when the paper will appear) is given in
Appendix~C.

The practice appears to have started in 2002, with a pair of papers discussing
the rivalry between students in the Lunar and Planetary Laboratory and
those in the Steward Observatory at the University of Arizona
\cite{arXiv0204013,arXiv0204041}.  These can't be claimed to be the
funniest examples, but they {\it were\/} first (and they're short)!\@
The next example was ``Cosmic Conspiracies'' in 2006 by me and Ali Frolop (more
of which later) and then the flood gates opened.  At this point there have
been at least 72 such papers, some much funnier than others (but I leave
readers to make up their own minds about the ranking).

An amusing fact: it can be hard to tell whether papers are meant to
be April Fools or not.  Without listing examples,
papers come out each year at the start of April that
appear so ridiculous that many people may regard them as jokes!\@
Temerity forbids me from listing any such cases.

\subsection{Ali Frolop}
I first collaborated with Dr.\ Frolop on a paper called ``Cosmic
Conspiracies'' in 2006 \cite{Frolop1}.
The inspiration was studying the cosmological parameters for sufficiently
long that a few apparent oddities had started to appear,
for example that $H_0 t_0$ is close to unity (with the accelerating and
decelerating phases more or less balancing each other).  It was Dr.~Frolop's
idea to submit a collection of these peculiarities
to the arXiv and not attempt to publish elsewhere.
But in fact, on further consideration there were even more near
coincidences to point out, leading to a semi-serious paper written with
Don Page and my student Ali Narimani, called ``Cosmic Mnemonics''
\cite{mnemonics} -- the motivation was now to have fun with numbers that
describe the Universe, giving others a toolkit of possibilities for
remembering various relations.  Being too technical for a popular magazine,
but too light-hearted for a journal, it was extremely hard to find a place for
such an article to get published.  In the end ``Physics in Canada'' gave our
paper a home.

With Dr.\ Frolop,
a  series of further papers followed, the topics covered being:
the appeal of multiple kinds of darkness \cite{Frolop2};
galaxies don't form at all, in fact they disappear \cite{Frolop3};
the CMB is {\it really\/} an inside-out star \cite{Frolop4};
we should abandon falsifiability {\it and\/} other cherished principles
as well \cite{Frolop5};
there are as many anomalies in the digits of $\pi$ as in the CMB sky
\cite{Frolop6};
there's a new kind of radio transient source that shows up before you look
for it \cite{Frolop7};
and normal logic doesn't apply to the search for life \cite{Frolop8}.
The ``Pi in the sky'' article is by far
the longest and contains the most serious message.  It's also the only paper
I'm aware of that has a word-search puzzle in it (and a rare example of a
1st April paper with a version~2 on the arXiv).

\section{Maxwell}
James Clerk Maxwell, as already stated, was a great joker.
As a youth in 1842 he wrote in a letter
``On Friday there was great fun with Hunt the
Gowk; we could believe nothing, for the clocks were all
`stopped', and everybody had a `hole in his jacket'\thinspace'' \cite{Gowk}.
``Hunt the gowk'' is the Scots phrase for pranks played on April Fool's Day
\cite{Poisson}.
In southern Scotland, where the young James Clerk grew up (before he was
``Maxwell''), it was also known
as ``hunt the dunse'', the word ``dunse'' being originally an epithet for
theologian and scholar John Duns Scotus, who lived in the 13th century.
This information is important to understand one of Maxwell's
biggest physics jokes.

In 1861 he started publishing a series of papers describing his new unified
theory of the electric and magnetic forces.  Maxwell realised that the
theory would be more self-consistent if there existed an additional kind of
field.  Initially there was no reason to believe in the existence of such a
field, but on 1st April 1861 Maxwell wrote to his older physics colleague
Michael Faraday, describing how he had evidence for this field, which he
named ``$D$'', after the letter labelling children who were being made a fool
of, or turned into a dunse on that day \cite{Dfield}.  Maxwell \cite{Daftie}
came to understand that the joke had backfired soon afterwards,
since he found that he actually needed
this additional component in his electromagnetic equations.  He
therefore dropped the ``dunse'' label and started to refer to it as the
``displacement field,'' $D$ \cite{Dunse}.  We now know it plays an
important role in completing Maxwell's equations, accounting for
the effects of free and bound charge within materials.  It's ironic that this
story all started with a prank.

\section{Moon hoax}
Through human history there has been speculation about the nature of the
Moon.  These conjectures included the famous astronomer William
Herschel stating in 1780 that there was a ``great
probability, not to say almost absolute certainty, of her being
inhabited'' \cite{HMoon1} and in 1795 he added that ``the analogies that have
been mentioned are fully sufficient to establish the high probability of
the moon's being inhabited like the Earth'' \cite{HMoon2}.  To make it clear
that conventional thinking was different back then, it may be worth noting that
Herschel also stated that ``we need not hesitate to admit
that the sun is richly stored with inhabitants'' \cite{WHerschel}.

Hard though it may be to accept, these were not jokes but serious suggestions,
albeit without any proof.
However, dramatic evidence of life on the Moon appeared in a series of articles
published in The Sun newspaper in New York in 1835 \cite{hoax}, apparently
based on new observations by William Herschel's son John.  These articles
discuss how forests, fields and beaches could be seen on the lunar surface
and, with a little more scrutiny, bisons and sheep,
as well as bipedal beavers, blue goats, unicorns and man-bats \cite{manbats}.
The articles caused a sensation at the time, with claims that, over the week
they appeared, the circulation of the newspaper increased dramatically.
Other newspapers in New York reprinted the stories, the original
publisher produced a pamphlet including the entire series, and there were
translations into many languages.  The story thus reached a very wide audience,
being an early example of the power of mass media.
It seems that a large fraction of people at the time genuinely
believed the hoax.  This would foreshadow the popular reaction to another
hoax a century later, this time using the medium of radio, namely
Orson Welles' ``War of the Worlds'' broadcast of 1938.

The Moon hoax was eventually debunked through several articles by journalists
and scientists questioning many details of the story.  The author was revealed
to be reporter Richard Adams Locke.  Locke claimed later that the story was
meant as a satire, attacking earlier works such as the 1824 paper
``Discovery of Many Distinct Traces of Lunar Inhabitants, Especially of One
of Their Colossal Buildings'' by Franz von Paula Gruithuisen (Professor of
Astronomy at Munich University) \cite{MoonCity} and the lunar-life beliefs
of Rev.\ Thomas Dick (who later calculated that the number of inhabitants of
the Solar System was 21,894,974,404,480 in total) \cite{factor}.
Undermining this declaration that it was all meant as a prank, Locke would
later try unsuccessfully to perpetrate another hoax, this time 
claiming to have the lost diary of the explorer Mungo Park \cite{MoonLanding}.

\section{Catching lions}
An influential piece of humour was published in 1938, called ``A contribution
to the mathematical theory of big game hunting'', written by H.\ P{\'e}tard
\cite{BigGame}.  Its subject matter may seem outdated, but I should reassure
you that the paper is about methods for capturing lions, rather than shooting
them \cite{Lions}; perhaps the author simply wanted to catch a lion in order
to take it to the Tower of London to be washed?  The author's name
was a pseudonym for mathematician Ralph P.\ Boas Jr.\ and some of his
colleagues (with the ``H'' being short for ``hoist with one's own'').
The paper describes 16 imaginative
methods for capturing a lion, and although the title says ``mathematical''
in fact four of the methods come from theoretical physics (including references
to one of Bethe's papers) and three from
experimental physics.  This paper led to additional methods of feline trapping
being later contributed by others, and it undoubtedly inspired further humorous
articles in physics as well as mathematics.  Boas used different
pseudonyms for other contributions,
including E.S. Pondiczery from the Royal Institute of Poldavia;
the original motivation was to use this author in a paper spoofing
extrasensory perception, so it could be signed ``E.S.P.\ R.I.P.'', but
unfortunately the paper never appeared \cite{NeverWere}.

\section{Candlestickmaker}
Subrahmanyan Chandrasekhar was one of the most highly-regarded astrophysicists
of his generation.  Originally from India, he spent most of his working life
in Chicago.  He received numerous awards, including the Nobel Prize for
Physics in 1983, and many things are named after him, including limits,
numbers, equations, functions, tensors, lemmas, asteroids and satellites!\@
He served as editor of the Astrophysical Journal (ApJ) for almost 20 years,
and had a reputation for reading all the papers submitted to the journal.
In 1957, Chandrasekhar's postdoc John Sykes (perhaps with the help of other
postdocs) wrote a parody paper called ``On the imperturbability of elevator
operators: LVII'', which claimed to be by S.\ Candlestickmaker from the
Institute for Studied Advances, Old Cardigan, Wales \cite{Candle}.

Sykes had come from Britain to join Chandrasekhar's group for a year to learn
about magnetohydrodynamics for the U.K. fusion project, but afterwards switched
to work as a physics translator and later dictionary editor.  He was able to
translate science papers in a large number of languages, and contributed to
the English version of the set of Landau \& Lifshitz volumes, for example.
He served as editor of the Concise Oxford English Dictionary and was a cryptic
crossword-solving champion par excellence (winning the annual Times
newspaper challenge no less than 10 times) \cite{Sykes}.

The ``elevator operators'' paper was submitted formally
to the ApJ.  The secretary, noting that it was probably
a joke, showed it hesitantly to Chandrasekhar -- but he was delighted with
the parody and insisted that it should be printed in the form of an ApJ
reprint.  This is the way that most people saw the paper at the time, and
it is said that several libraries bound it along with the regular journal.
The great man was so pleased with how well it captured his style that he
would recommend it to new students as a template for how to write a paper
\cite{Osterbrock}!

The specific paper being lampooned was probably ``The Instability of a
Layer of Fluid Heated Below and Subject to the Simultaneous Action of a
Magnetic Field and Rotation. II'' \cite{Chandra}.  A direct comparison is
needed to fully appreciate the in-jokes.  This ``Candlestickmaker'' paper
represented a seminal moment in the history of parodies, setting a high
bar for all others to follow.

\section{Thiotimoline}
In 1948 the science-fiction writer Isaac Asimov published a spoof article on
``The Endochronic Properties of Resublimated Thiotimoline'' \cite{thiotimoline}.
The inspiration for the article was Asimov watching substances dissolve almost
before they hit the surface of a liquid, during experiments carried out for his
chemistry doctorate at Columbia University.  The idea in the
paper was that a special compound had been discovered that dissolved
{\it before\/} making contact with water.  This could then lead to effects
that messed with causality (so although this starts with chemistry, now it
enters the realm of physics) and hence time travel.  As Asimov later
explained, the publication appeared shortly before his
doctoral defence with his own name rather than the planned pseudonym,
and he became concerned that the examiners might be annoyed that
he wasn't taking chemistry seriously.  However, at the end of the meeting,
one of them had a
last question, which was ``Can you tell us about the endochronic properties of
resublimated thiotimoline, {\it Dr.}\ Asimov'' and he knew he had passed!

Asimov followed his original paper with another on ``The Micropsychiatric
Applications of Thiotimoline'' \cite{Thio2} and a third on ``Thiotimoline and
the Space Age'' \cite{Thio3}, which includes a discussion of the ``Heisenberg
failure'', where it seems to be impossible to avoid adding water to the
substance after it had dissolved.  Lastly Asimov wrote ``Thiotimoline to the
Stars'' \cite{Thio4}, where he mocks his own name as ``Azimuth'' and
``Asymptote'' and includes a discussion of how endochronicity can be used to
power starships.  Several other science-fiction authors have also made mention
of thiotimoline.  It has therefore taken on an existence beyond
that of its inventor, joining legendary substances such as administratium,
neutronium, red mercury, unobtanium, etc., as well as devices like the
flux capacitor, the turboencabulator and write-only memory.

\section{Interstellar economics}
Another influential joke paper is ``The Theory of Interstellar Trade'',
written by economist Paul Krugman in 1978 \cite{interstellar}; it was
published in 2010, a couple of years after Krugman was awarded the
Nobel Prize in Economics.  The paper describes how trade,
interest and arbitrage might work when journeys could
take centuries and when the time passing in the frame of the traveller could
be different than the time passing on Earth; hence the
paper uses the ideas of special relativity, with the application of general
relativity ``left as an exercise for interested readers because the author does
not understand the theory'' \cite{GR}.  The paper states that it ``is a serious
analysis of a ridiculous subject, which is of course the opposite of what is
usual in economics'', an attitude that has influenced later spoof papers.
Several other studies of space trade have followed, and although some,
like Krugman's original paper, have their tongues firmly in their cheeks
\cite{tax}, it's unclear if that's true of all of them \cite{finance}.

\section{Gravity}
The ``Jupiter Effect'' was a suggestion that a planetary alignment in 1982,
when eight of the nine known planets (excluding Pluto) would all be on the
same side of the Sun, was going to lead to catastrophes on Earth.
The sensationalist book of the same name \cite{Jupiter},
written in 1974 by John Gribbin and Steve Plagemann,
caused extremely skeptical reactions from most scientists.  And in fact Gribbin
and Plagemann published a sequel in 1982, pointing out why it was obvious all
along that the effects on the Earth would be entirely negligible.
One wonders whether this was all some sort of deliberate hoax; however, based
on Gribbin saying later that he regretted having anything to do with this,
it would appear to have been over-enthusiasm in writing a popular book
rather than a prank.  Nevertheless, there {\it was\/} a later hoax associated
with this book, since it was the essential inspiration for a
trick perpetrated on 1st April 1976.  TV-astronomer Patrick
Moore claimed that at 9:47$\,$p.m.\ a conjunction of Jupiter and Pluto would
take place, resulting in a decrease in gravity on Earth, so that if people
jumped in the air at that precise time they would experience a form of
levitation.  Inevitably, many people reported that they felt floating
sensations at the appointed time \cite{Moore}.

\section{The Sokal Affair}
Sometimes there is a serious purpose behind a scientific prank.  Such was the
case with one of the most famous spoof papers in physics, through something
usually referred to as ``The Sokal Affair'' \cite{Sokal}.
The motivation was a feeling
among some scientists that postmodernists had gone too far in pushing their
agenda of knowledge being entirely based on social conditioning -- they
had become anti-science.  Moreover, there was a suggestion that some relevant
journals had a low quality threshold \cite{GrossLevitt}.
So, in 1994 physicist Alan Sokal wrote the spoof paper
``Transgressing the Boundaries: Towards a Transformative Hermeneutics of
Quantum Gravity'' \cite{SokalFootnotes},
which was accepted for publication in the journal ``Social Text'' in 1996
\cite{SocialText}.  
The main tenet of the paper was that physicists had for too long
ignored the views of deconstructionists and should free themselves from
the restrictions of things like mathematics.

Here's a snippet from the paper, where Sokal responds to a comment from
deconstructionist Jacques Derrida about general relativity:
``In mathematical terms, Derrida's observation relates to the invariance of
the Einstein field equation $G_{\mu\nu}=8\pi G T_{\mu\nu}$ under nonlinear
space-time diffeomorphisms (self-mappings of the space-time manifold which are
infinitely differentiable but not necessarily analytic). The key point is that
this invariance group `acts transitively': this means that any space-time
point, if it exists at all, can be transformed into any other. In this way the
infinite-dimensional invariance group erodes the distinction between observer
and observed; the $\pi$ of Euclid and the $G$ of Newton, formerly thought to be
constant and universal, are now perceived in their ineluctable historicity; and
the putative observer becomes fatally de-centered, disconnected from any
epistemic link to a space-time point that can no longer be defined by geometry
alone.''

Sokal quickly confessed that the article was a deliberate spoof,
containing largely nonsense couched in technical-sounding physics jargon.  Some
postmodernists were upset that they had been fooled into taking the
article seriously; they deemed it as unfair because they were not experts
in physics, despite the fact that the text contains many glaring clues to its
insincerity, and not excusing the fact that the journal failed to ask for the
opinions of any specialist reviewer before publishing \cite{SokalBeyond}.
Perhaps most amusingly, other
postmodernists argued that it was irrelevant that the article was intended as
a joke, since it stood as an academic treatise on its own merits!

\section{More hoaxes}
Chemistry provides an example of a deliberate hoax from a publication in 1944,
``Toxicological Significance of Laevorotatory Ice Crystals''
\cite{Laevorotatory}.  The paper
purports to be about how left-handed ice crystals were poisonous; but in
fact the paper was a deliberate ``sting'' perpetrated on the editors of the
journal ``The Analyst'' after some dispute with them.
The paper was subsequently withdrawn and not indexed, but is now easy to track
down.

One of the scourges of modern academia is the proliferation of predatory
journals, i.e.\ new journals of low quality and little in the way of peer
review.  There have been several attempts to show that some of these journals
will publish just about anything, provided that they get their fee.

An example is the 2017 paper
``Mitochondria: Structure, Function and Clinical Relevance''
by Lucas McGeorge and Annette Kin, which substitutes ``midichlorians'' for
``mitochondria'' in the text -- these are the microscopic creatures introduced
in later Star Wars films to explain ``the Force''.  The paper was accepted by
four journals, including the Austin Journal of Pharmacology \& Therapeutics
\cite{midichlorians}; two of these journals later removed the paper.  
This Star-Wars-inspired spoof was quickly followed by another motivated by Star
Trek.  ``Rapid genetic and developmental morphological change following
extreme celerity'' by Paris, Kim, Torres, Ocampa, Janeway \& Zimmerman
\cite{celerity} was based on the Star Trek Voyager episode ``Threshold''
\cite{threshold}.  This paper was sent to 10 apparently predatory journals,
four of which accepted it, and one published it (although the paper was later
removed from the publication website).
Unfortunately this has all had zero impact on the spread of such journals.

In 1897 the ``Indiana Pi Bill'' was a famous attempt to effectively
legislate that $\pi=3.2$ (or perhaps $3.23$ or maybe even $4$, it's hard to
tell).  Variants of this story have done the rounds for decades, often
involving the value 3 and locations in different states.
In 1998 physicist Mark Boslough was able to
exploit this confusion by publishing an April Fool's Day joke as a satirical
attack on the pro-creationist stance of New Mexico's state legislature
\cite{NMSR}.
Boslough wrote that Alabama had voted to set $\pi=3$ and people starting
calling their representatives to complain, leading to the prank being revealed.

\section{Monopoles}
On Valentine's Day in 1982, an experiment designed by Blas Cabrera recorded an
event that had all the characteristics of the hypothesised magnetic
monopole \cite{Cabrera}.  This
caused a great deal of excitement at the time, but no further events were
seen and later experimental results placed very stringent limits on the flux
of monopoles.  One problem is, since it was a weekend, no one was scheduled
to be in the lab that day.  It has been suggested that the apparent
event may have been a prank
played by a student \cite{monopole}; however, no one has ever owned up!\@
The possibility of it being a prank is interesting because,
just a year before, Sidney Coleman, in some published physics lectures
\cite{Coleman}, had introduced the ``monopole hoax'' as a way for theorists
to fool experimentalists that they'd seen a monopole using ``a very long, very
thin solenoid \dots many miles long'' having one end in the laboratory and
the other very far away.  Could this light-hearted suggestion
have been the inspiration for a practical joke?

\section{Other spoof papers}
An early example of a scientific lampoon
from the Journal of Irreproducible Results in 1956 was
``Theoretical zipperdynamics'' \cite{Zipper}.  This paper described the
quantized nature of the position of a zipper, the unobservability of
``zipperbewegung'' and attempts to solve the semi-infinite zipper using the
Schroedzipper equation.  It included references to several earlier papers,
including one by H.~Quantum on zipper theory, ``which is incidentally
applicable to such minor Problems as Black Body Radiation, Atomic
Spectroscopy, Chemical Binding and Liquid Helium''.

Although editors tend not to be known for their senses of humour,
occasionally, mainstream journals will include tongue-in-cheek contributions;
such was the case with the 1970 paper in Science called ``Properties and
Composition of Lunar Materials'' by Schreiber \& Anderson \cite{Cheese}.  This
paper presents the results of experiments to measure the sound speed in various
substances, finding that the terrestrial materials that most closely match
the Moon are various types of cheese.

One physics paper published in the Worm Runner's Digest was 1972's
``A theory of ghosts'' by D.A. Wright \cite{Ghosts1}.  The article uses quantum
mechanics, relativity and other bits of physics to describe how ghosts can
penetrate walls, move quickly and be observed at low light levels.  Additionally
the author questions whether ghosts are fermions or bosons and speculates that
they could be the source of the cosmic microwave background.

In 1979 a preprint by De R{\'u}jula, Ellis, Petronzio, Preparata \& Scott
\cite{Hole1},
based on a dramatic performance at the Erice Particle Physics School, was
entitled ``Can one tell QCD from a hole in the ground?''\@
Ellis followed the same theme in 1980 with ``Can One Tell Technicolor from
a Hole in the Ground?''\! \cite{Hole2}

A preprint from 1980 by ``Doctor'' Wisecracker \cite{Stuff} was entitled
``Is the Universe full of stuff?''\@
It includes statements such as ``In the standard model the
cosmos starts as a huge banana stuffed with quantum foam''.
With the prominence of the concepts of supersymmetry and
superstrings, G.\ Wow-mann wrote a follow-up called ``Superduperstuff in the
Universe'' \cite{Superduperstuff}.
This was based on the concept of ``superconducting,
supercolliding, supersymmetric, superstringy superstuff'' with which
``any phenomenon can be explained by a theorist of arbitrary skill''.
The author of these parodies was particle astrophysicist
Craig Hogan, during his time as a graduate student and postdoc.
The superabundance of the prefix ``super-'' in physics was developed
by others in a more recent paper on ``Superfluous Physics'' \cite{superfluous}.

Another unpublished preprint came out in April 1983, ``Monte Carlo Simulation
of a Realistic Unified Gauge Theory'', by Alan Chodos and Jeffrey
Rabin \cite{Chodos}.  It introduced the idea of the ``Grassmann Chip'', which
could store and compute with anti-commuting numbers (and has later been
discussed in more serious papers).  Theorist Joseph Lykken wrote
``Observation of Warm Nuclear Fusion in Condensed Soup'' in 1989 \cite{Soup},
in response to the contemporary claims of cold fusion.

String theorist Warren Siegel has published a series of paper parodies
as preprints \cite{siegel}, 22 at the latest count!\@
This makes him one of the most
prolific of the physics paper spoofers.  The series began with the 1983 paper
``Stuperspace'', purportedly by V.\ Gates, Empty Kangaroo,
M.\ Roachcock \& W.C.\ Gall, and later published in Physica D
\cite{stuperspace}.  These authors have remained together as a team,
although curiously from 1993 we find that Dr.\ Kangaroo has dropped down the
priority list relative to Dr.\ Roachcock.  Readers of these papers will notice
that their authors are very fond of footnotes and the overuse of elaborate
typography.  Additionally, one paper refers to the ``Newton-Witten''
equation $F=ma$, which gained a certain infamy.
Whether consciously or not, these papers seem
to be inspired by the earlier ``Candlestickmaker'' parody.

``Script an Astronomer, Then Reach for the Stars'' by Eric Schulman appeared in
AIR in 1999.  It describes the positive correlation between the quality of
movies and the number of characters in them who are astronomers \cite{script}.
Along with several joke articles and poems \cite{Schulman},
Schulman also contributed the parodies ``How to Write a Scientific Paper''
\cite{HowTo1}, ``How to Write a Scientific Research Report'' \cite{HowTo2} and
``How to Write a Clear Research Report'' \cite{HowTo3}.  The last two,
written by Schulman with the help of his partner and their daughter,
describe ``the stacking properties of toroids that reflect radiation in the
1.8 to 2.8\,eV energy range''.  Additionally
``The insulating properties of materials'' describes the finding that
``newsprint has superior thermal insulation properties when compared to
corrugated fiberboard or air cellular cushioning material'' \cite{insulating}.

A paper posted to the arXiv in 1999, called ``The Effects of Moore's Law and
Slacking on Large Computations'' showed that, based on the rate at which
computational power is growing, it's better to wait and carry out your
calculations later \cite{slacking}.  The authors, Gottbrath et al., were
students at Steward Observatory.

Continuing the tradition of using animals in spoof papers, ``The Violation of
Bell Inequalities in the Macroworld'' written by Aerts et al.\ in 2000,
contains a section involving cats with bells tied around their necks
\cite{Aerts}.
It is unclear how much of the rest of the paper has its tongue in its cheek.

Physicist Donald Simanek (and his very close colleague Ken Amis)
has written several science parodies, some intended
to test critical thinking abilities in students.  These include ``A  New
Theory of Dark Matter'', ``The Age of the Universe is a Function of Time'',
``A Deductive Proof of Newton's Third Law'' and ``Toward a New Theory of
Gravitation'' \cite{Simanek}.

In 2009 physics graduate student Ben Tippett
wrote ``A Unified theory of Superman's Powers'' \cite{Superman}.  The abilities
of Kal-El are usually explained to arise from Earth's gravity being weaker than
the planet Krypton's, along with our Sun being yellower than Krypton's
star, but Tippett argues that these explanations make little sense and that it
would be much simpler to assume that Superman could just manipulate inertial
mass.  There are follow-up papers on Spiderman's super-powers \cite{Spiderman},
a lost city from a story by H.P. Lovecraft \cite{Rlyeh} and Dr.\ Who's TARDIS
\cite{TARDIS}.

``A Simple Model of the Evolution of Simple Models of Evolution'' by
Shalizi \& Tozier \cite{adaporg9910002} is a spoof paper making a serious
point.  It's about the tendency of some physicists to write papers modelling
evolution as a statistical-mechanics problem, without really knowing any
biology.

Let me give one last example that I dimly recall hearing in my early days as a
graduate student.  Apparently there was a conference where a senior
astronomer presented results showing that many stars had rings around them,
with the angular size of the rings appearing to be inversely proportional to
the distance of the star -- in other words there were structures of fixed
physical size around many stars.  Hence this couldn't be an optical
effect and had all the appearances of being artificial constructions, something
like Dyson spheres.  Despite this seeming to be the most amazing discovery
ever made, most people listening to the talk showed no reaction, and it
transpired afterwards that the astronomer had been pranked by his own
graduate student!\@  Unfortunately I've been unable to uncover where I heard
this story, or to find out any more information.  So perhaps I just made
this up? \cite{rings}

\section{Backfires}
Sometimes comments that are intended to be satirical end up having the opposite
consequences.  There are two famous examples.  Firstly, ``Schr{\"o}dinger's
cat'' was originally a thought-experiment devised by Erwin
Schr{\"o}dinger in 1935 to
ridicule the Copenhagen interpretation of quantum mechanics.  
As he wrote: ``One can even set up quite ridiculous cases. A cat is penned up
in a steel chamber, along with the following device \dots'' \cite{Schrodinger}.
However, this thought experiment
has come to be considered as a serious manifestation of the principle of
superposition of states, losing the negative connotations that were its
initial intent.  The image of the dead/alive cat has also grown in prominence
in popular accounts of quantum mechanics, probably being the focus of more
jokes (cartoons, t-shirts, etc.)\ than any other topic in physics;
most of these jokes completely
misinterpret the idea, as well as missing the fact that the situation was
{\it itself\/} meant as a joke \cite{CatJokes}.

The second example is the naming of the ``Big Bang'' model for the early
history of the Universe.  The phrase was coined by Fred Hoyle in a radio
broadcast in 1949.  Hoyle was a proponent of the rival ``steady state'' theory,
which posited that the Universe has always looked essentially the same, and he
argued against a picture where time had a beginning.  Hoyle appeared to use
the term ``Big Bang'' pejoratively, specifically saying that the theories he
didn't like ``were based on the hypothesis that all the matter in the universe
was created in one big bang at a particular time in the remote past''
\cite{BigBang}.  The debate between proponents of the ``big bang'' and the
``steady state'' went on for a couple of decades, as summarised in a poem by
Barbara Gamow (wife of George \cite{Blunder}), including these lines:
\par
{\sl Said Hoyle, ``You quote}
\par
{\sl Lema{\^\i}tre, I note,}
\par
{\sl And Gamow. Well, forget them!}
\par
{\sl That errant gang}
\par
{\sl And their Big Bang --}
\par
{\sl Why aid them and abet them?}
\par
{\sl You see, my friend,}
\par
{\sl It has no end}
\par
{\sl And there was no beginning}
\par
{\sl And Bondi, Gold,}
\par
{\sl And I will hold}
\par
{\sl Until our hair is thinning!''}
\par\noindent
It is therefore ironic that the phrase `Big Bang'
eventually became attached to the theory
that Hoyle scoffed at.  And in a further level of irony, the name is
generally detested by cosmologists because it conjures an image of an initial
explosion (the $t=0$ instant generally being considered to be {\it outside\/}
the purview of the model) at a specific place.

The moral here is, if you're going to come up with a good sound bite to
lampoon something, then don't be surprised if it comes back to bite you!
\cite{biteyou}

\section{Ephemera}
When print media was the norm, comic items related to science would appear in
``grey material'' venues like newsletters, which may not have been
well archived.
An example is ``Physikalisches Lied'' by Molly Kule, consisting of what looks
like a piece of music, but is crammed with physics-based visual jokes and puns.
This seems to have come from Princeton in about 1942, and was preserved in
``More Random Walks in Science'' \cite{random2}.

Before there were ``e-prints'' there were ``pre-prints'' \cite{preprints}.
The eprint archive started in the early 1990s in various sub-fields of
physics.  Before that time it was common practice for people to prepare
preprints that were circulated to major institutions around the world, so
that the results could be disseminated during the delay before journal
publication.  Joke papers would occasionally be added to the bundle of
such preprints being mailed out.  Unfortunately such contributions
are therefore impermanent, and may only exist in filing cabinets.
Several examples have already been mentioned above, but
there are probably quite a few spoof preprints out there yet to be unearthed.

Another example of a set of ephemera are the joke emails that were
used to exchange casual information before the rise of the
internet.  I can remember one in the form of a ``chain email'', encouraging
the reader to cite specific papers, add a paper of their own to the list,
and send the same email to ten colleagues -- or else bad things would
happen.  There was also the ``cartoon laws of physics'', which came with
several amendments \cite{cartoon}.
Other examples included spoof versions of referee reports and
how to deal with them, proofs of the impossibility of Santa, and where to
order items like spherical cows, frictionless planes and massless springs.
Some items are collected in repositories from usenet newsgroups
\cite{Verhagen}, but
one wonders whether anyone has a full collection of these
emails from the 1970s and 1980s.  Going back even earlier, there will be
similar content in actual letters on pieces of paper!

One story, repeated in many forms, is based on ways of using a barometer to
measure the height of a building.  This originated in about 1960, written by
physicist and educator Alexander Calandra \cite{Calandra}.  It has been
reprinted and embellished many times, to the extent that it became an urban
legend.

Posters at conferences are also largely unrecorded.
In about 1986, ``A New and Definitive Meta-Cosmology Theory''
was a flow chart created by Lauer, Statler, Ryden \& Weinberg, who were
then Princeton graduate students.  It describes how the discovery of a new
particle can be developed into a
cosmological model, all the arrows ultimately leading to $\Omega_0=1$,
which was the conventional wisdom among theorists at that time.
Additionally, a one-off board game called ``Galaxy Formation'' was created
by David (and Lisa) Weinberg in 1987 and played at some conferences.

\section{Paper sections}
We have already seen many cases where a whole paper is a
lampoon of some aspect of physics or astronomy.  But there are many more
instances where the joke occurs only in a small part of the publication.
So let us now go through the various segments of a paper, giving some examples
of whimsy for each of these parts.

\subsection{Pre-publication}
There are several steps in preparing a paper for publication.  Perhaps first
comes the inspiration.  Something that might help is this quatrain:
\par
{\sl God grant that no one else has done}
\par
{\sl The work I want to do,}
\par
{\sl And give me wit to write it up}
\par
{\sl In decent English too.}
\par\noindent
This was submitted as part of a competition in 1962 for a chemistry version of
``The Fisherman's Prayer'' \cite{Godgrant}, with the author recorded as
``Ricardo''.

Another important part of the research process is applying for funding.
wrote ``Creation of the Universe: a modest proposal''
as a parody of a grant proposal, seeking additional finances to correct
some of the flaws in the Universe that were created following a previous
round of funding \cite{Creation}.

\subsection{Authors}
Probably the best-known author list created for humorous impact occurs
in the 1948 paper by Alpher, Bethe \& Gamow on ``The Origin of the Chemical
Elements'' \cite{ABG}.  This remains a seminal study in the history of
ideas for the formation of the light elements.
The work was done by graduate student Ralph Alpher, along with his
supervisor George Gamow.  It was Gamow's idea to include his friend Hans
Bethe in the author list \cite{Bethe}, partly because he learned that
the paper would appear on 1st April \cite{Delter}.
Alpher apparently did not appreciate the joke \cite{Alpher}.
However, despite the fact that Alpher thought the addition of a
non-participatory senior scientist would somehow lessen his perceived
contribution, or lead to the paper being taken less seriously, in fact
it probably gained prominence through being
known as the ``$\alpha\beta\gamma$'' paper \cite{Fame}.

Greenberg, Greenberger \& Greenbergest posted a paper to hep-ph in 1993
\cite{Green}.
In 2011 a paper appeared on 1st April on patterns
in the cosmic microwave background, written by Zuntz, Zibin, Zunckel \&
Zwart \cite{ZZZZ}; a related group also showed that authors near
the end of the alphabet get fewer citations but write better papers
\cite{arXiv10036064}.  Although outside physics, it's worth noting that
in 2014 Goodman, Goodman, Goodman \& Goodman studied papers by authors sharing
a family name in ``A Few Goodmen: Surname-Sharing Economist Coauthors''
\cite{Goodmen}.

A paper published in 1989, called ``The small-scale autocorrelation function of
the X-ray background'', was written by Xavier Barcons and Andy C. Fabian.
``What's funny about that?'' you might ask.  The answer is given in a footnote
on the first page of the paper: ``The small-scale ACF of the XB by XB and
ACF'' \cite{XBACF}.  
Several people with the surname ``Moon'' have written papers about the Moon,
while Wolfgang Wall has modelled walls in fluid mechanics.
There are surely other examples of people working on topics that are
apposite to their names.

In 1989 spectroscopist Peter Hollins found he had a set of name-appropriate
students in his group and hence a paper by Quick, Brown, Fox \& Hollins
was born \cite{QuickBrown}.
An article on ultrashort laser pulses, published in ``Optics \& Photonics
News'' in 1990, claimed to have reached the limit of a zero-width pulse,
and that in future pulses of negative width would be possible.  The
authors were Knox, Knox, Hoose \& Zare \cite{KnockKnock}.  The authors' names
are real, but one suspects they got together merely for the purpose of
writing this spoof.  A paper from 1992 on
${}^{13}{\rm C}$--${}^{13}{\rm C}$ couplings was authored by Bax, Max \& Zax
(the first author being a biophysicist) \cite{Bax}.
Also in 1992, D'Eath \& Payne co-authored three papers together about
gravitational waves from black holes \cite{DEath}.

The physicist Alois Kabelschacht first appeared as a colleague acknowledged
for help, and then in 1978 was promoted to the status of co-author for a theory
paper \cite{Kabelschacht}.  He has now appeared on three other papers
\cite{Alois}, including two as an experimentalist; however, the name
is a joke deriving from the nameplate on what looked like an office door
at the Max-Planck-Institut f{\"u}r Physik in Munich, but
was just the ``cable shaft''.

J.J.~Charfman is an astronomer of legendary status from the Steward
Observatory.  The first
paper they wrote was about boron sulphide, which terrestrially occurs as
B$_2$S$_3$ but in the interstellar medium apparently exists in the form of
just one B atom attached to one S atom \cite{BS}.  The same author's name
appears four more times; it is unclear where the name came from
\cite{Charfman}.

T.I.A.~Fudge was added as co-author to a paper on modelling Bose-Einstein
condensates in 2002.  This fictitious author came from a confession
that a certain coefficient
(unexplained at the time) was a ``fudge factor'', with the initials standing
for ``This Is A'' \cite{Fudge}.
Physicist Jack H. Hetherington attached the name of his cat Chester, through
the alias ``F.D.C. Willard'', to one of his papers in 1975 \cite{cat}.
Apparently this originated in a debate with a
colleague over whether the Physical Review would reject the paper for using
the first person plural in a manuscript
with just a single author -- a debate he resolved by adding a bogus
co-author.  ``F.D.C.'' stood for ``Felis Domesticus, Chester'' with Willard
being the name of Chester's father.  Willard later wrote a single-author
article in French for a popular science magazine \cite{Willard}.
The practice of adding a pet as co-author was repeated (with the same
``third person'' reasoning) by immunologist Polly Matzinger in 1978 \cite{dog}.

Andre Geim, famous as the only person to have won both a Nobel Prize
\cite{nobel} and an Ig Nobel Prize \cite{ignobel}, wrote a paper with his
hamster \cite{hamster}, H.A.M.S. ter Tisha.  It is unclear what the initials
stand for, but clearly Geim went to less effort to hide the fact that
his co-author was a pet than the trend started by Hetherington.

A.~Aardvark is recorded as a co-author on several abstracts and papers; one
has to be suspicious that the name was added for alphabet-inspired reasons
\cite{Aardvark}.  Continuing the animal theme,
Tycho Brahe, known as something of an eccentric,
had a pet elk \cite{elk}; no doubt, if he had lived a few centuries later,
he would have included it as a co-author on some of his papers.
Sadly the elk died falling down the castle stairs after getting drunk.

Speaking of bogus authors,
let me come clean and confess that the name ``Ali Frolop'' is made up.  This
had an unintended consequence when Ali Narimani became my graduate student a
few years later.  Several people noticed I'd been publishing with
Mr.\ (now Dr.) Narimani, and asked me what that name could be an anagram of!
I had to respond by assuring them that Ali Narimani is in fact a real person.
But this got me thinking that it would be fun to collaborate with both of
the Alis in my research life.  So when Ali N.\ came to me with an idea for
a spoof paper (picking up on something he'd read that Sean Carroll had
written), the project was started.  It then seemed obvious that we should
also recruit Andrei Frolov, a colleague from our neighbouring university,
allowing us to concoct the following joke author list:
``Ali Frolop, Ali \& Frolov'' \cite{Frolop5}.
From my perspective this author list became the main reason to write the
paper!\@  Andrei, like Hans Bethe before him, played no role in writing the
paper, but agreed to have his name added.  However, he wanted an assurance that
``nothing bad would ever happen to him'' as a consequence of being
associated with our joke.
Within a week the paper was completed and ready to be submitted to the arXiv
at the appropriate time.  However, there was no way to put ``D.\ Scott'' among
the authors while preserving any humorous impact, so I omitted my name from
the author field on the upload page.  This led to the paper bouncing back
from the arXiv, with a message that third-party submissions were not
allowed.  I then explained in a note that I {\it was\/} in fact an author, and
therefore added my name to the author field for the resubmission.
This led the site moderator to withdraw
submission privileges for us on the basis that we were trying to subvert the
arXiv's policies!\@  Andre wasn't happy.  And it took 24 hours to find someone
in a position to sort this out for us, which explains why the paper appeared a
day later than intended.  On the positive side, this story
may be funnier than the paper itself!

As well as Frolop, Kabelschacht, Charfman, etc.,
there are plenty of examples of scientists of
possibly legendary status.  Monsieur Litre was an early instrument builder,
while Konrad Finagle was the inventor of the fudge factor, as well as several
other innovations (as described by Donald Simanek).  Then of course there are
other physicists who have appeared in various places,
such as Dr.\ Arroway, Dr.\ Banner, Doc.\ Brown, Dr.\ Brundle,
Prof.\ Calculus, Dr.\ Koothrappali, Dr.\ Manhattan and Dr.\ Octopus.

To finish this section, I feel compelled to acknowledge the issue of increasing
numbers of authors on papers from large collaborations, undermining what
most of us thought we understood by the word ``author''.  The current record
holder is the 5{,}154 names listed on the 2015 Higgs boson paper from the two
combined experimental teams at the Large Hadron Collider \cite{LHC}.  This
paper, including references, ends on page~9, with the listing of authors and
institutions stretching it out to page~33.  At the other extreme there are
still plenty of single-author papers, including some with very short names
\cite{I}.

\subsection{Addresses}
In parodies of papers the authors are often listed at spurious institutions.
The ``Candlestickmaker'' paper gives the author's address as ``Institute for
Studied Advances'', the paper on zipperdynamics came from ``The Weizipmann
Inziptute'' and the origin for the ``Superduperstuff'' paper was
``Institute for Innerspace/Outerspace Interfarce''.
In one paper John Ellis gave his address as ``British Airways''.

\subsection{Titles}
Joke titles are quite common, although some of the older journals tend to
frown upon the practice.  Hence there are many cases where the arXiv posting
has an amusing title that has been changed to something much more boring for
the journal publication.  Let me give a few examples here, with a somewhat
longer list being provided in Appendix~D \cite{Medical}.

Although the use of humorous titles may seem like a fairly recent phenomenon,
there are some earlier examples.  For instance,
``Deuteronomy. Synthesis of Deuterons and the
Light Nuclei during the Early History of the Solar System'' was published by
Fowler, Greenstein \& Hoyle in 1961 \cite{Deut}.

Different topics within physics and astronomy have varying levels of zest for
jokey titles.
Amusing and outlandish titles became quite popular within string theory, 
and there are many examples, like ``10=6+4'' \cite{arXiv9908205} and
``Escape from the Menace of the Giant Wormholes'' \cite{ColemanLee}.
Within astrophysics, black hole theorists and cosmologists seem more
enthusiastic about joke titles than researchers in most other areas.

``Velocity dispersions in a cluster of stars'' by Eriksen, Kristiansen,
Langangen \& Wehus \cite{Bolt} is a clever title.  One might expect that this
is about the statistics of motion in a globular cluster, say.  But the
subtitle gives the game away: ``How fast could Usain Bolt have run?''\  This
is in fact a statistical study of frames from film of the famous race where the
sprinter appeared to slow down at the end.  Unfortunately the journal changed
the title to something much more prosaic.

There are a large number of titles based on movies, particularly from science
fiction.  Plays on the names of Star Wars films are especially popular
\cite{starwars} -- ``strikes back'' seems to be a phenomenon that occurs in
many branches of physics.  Similarly, there are a lot of ``one rings'' doing
something to ``them all''.  Shakespeare-inspired titles are also common, with
their ``Much Ados'', their ``All's Wells'' and their ``To Bes''.

Newspapers and magazines have more sensational headlines than we're used to
on the front pages of scientific papers.  The ``News and Views'' section in
Nature is somewhere between a magazine and a journal, and hence is a good
source for deliberately humorous titles, e.g.\ ``A new twist for cosmic
strings'' \cite{Twist}, ``Evading the zone of avoidance'' \cite{Evading},
``Goings on between the stars'' \cite{Goings},
``In search of the halo grail'' \cite{Grail} and
``White dwarfs sing the blues'' \cite{Blues}.

From my own papers, I'm particularly proud of getting
``Boomerang returns unexpectedly'' \cite{boomerang} accepted by the ApJ!\@
The explanation was that the paper interpreted surprising results from the
cosmic microwave background (CMB) experiment called ``Boomerang''.
The paper ``What have we already learned from the CMB?'' \cite{Romans}
started with a quote from Monty Python's ``Life of Brian'', while the short
title used in the page headings was ``What has the CMB ever done for us?''
Then ``Cosmological Difficulties with Modified Newtonian Dynamics''
\cite{MOND} had the subtitle ``La Fin du MOND?''
Another good title is ``Resolving the Radio Source Background: Deeper
Understanding through Confusion'' \cite{Condon}.  The only Planck Collaboration
paper with a less-than-serious name was ``Planck 2013 results. XXVII. Doppler
boosting of the CMB: Eppur si muove'' (named for the phrase said to have been
spoken by Galileo after he was forced to recant) \cite{Eppur}.
``Evaporating evidence
for Hawking points in the CMB'' \cite{Jow} was changed by the journal to
``Re-evaluating evidence for Hawking points in the CMB''.  Lastly, a
semi-serious overview of the history of ideas in the topic of galaxy formation
was entitled ``The evolution of galaxy formation'' \cite{EoGF}.

Does having a funny title actually help?
Interestingly, a serious study of whether adding an amusing title increases
the number of citations actually found a negative (although weak)
effect \cite{Sagi}.  A later study showed that papers with funny titles
tend to get more downloads, but not more citations \cite{Subotic}.

In addition to papers, conferences often have amusing titles.
One of the best may be ``TANGO in PARIS'', which was ``Testing Astroparticle
with the New GeV/TeV Observations Positrons And electRons: Identifying the
Sources'', which took place in 2009 \cite{TANGO}.
There was a conference in 2015 called ``Mocking the Universe'', but
disappointingly it turned out to be about numerical simulations, rather than
cosmological humour.

\subsection{Abstracts}
The paper ``Chern numbers, quaternions, and Berry's phases in Fermi systems''
by Avron, Sadun, Segert \& Simon \cite{Avron} has the following brief
abstract: ``Yes, but some parts are reasonably concrete''.  ``Are Magnetic
Dips Necessary for Prominence Formation?'' by Karpen et al.\ \cite{Karpen}
has the more perfunctory abstract ``The short answer: No''.  However,
the 2011 paper
``Can apparent superluminal neutrino speeds be explained as a quantum weak
measurement?'' has the even briefer abstract ``Probably not'' \cite{Berry},
which likely has the record as the shortest example in physics.

For Max Tegmark's 1996 ApJ paper on pixelizing the sphere, the abstract was
entirely in rhyming couplets \cite{Tegmark}.
Ben-David \& Sattath wrote an abstract based on the fairytale ``The Fisherman
and His Wife'' to introduce their 2017 paper on quantum cryptography
\cite{BenOr}.  A review of ideas concerning the origin of ultra-high energy
cosmic rays, written by J{\"o}rg Rachen in 2019, has an abstract inspired by
the original Communist Party manifesto of Marx \& Engels \cite{UHECR}.
Robert J. Nemiroff (co-founder of the APOD site) collected several items on
his ``Comedy of Science'' page, including joke versions of an abstract,
an erratum and an acknowledgements section \cite{RJN}.

A generator of fake abstracts (and titles) from high-energy physics is
provided at the snarXiv website \cite{snarxiv}.  The site was developed by
David Simmons-Duffin, who also created the ``arXiv versus snarXiv'' game.
In computer science the ``SCIgen'' site allows you to create whole papers,
including figures and references.

\subsection{Introductions}
In the reference work
``Atomic Transition Probabilities: Hydrogen through Neon'' \cite{HthroughN},
the fluorine section contains a statement that ``since we expect that this
introduction will share the fate of most introductions (namely be ignored)
\dots we might as well give the few readers of this introduction some good
advice:
\par
{\sl If there is no other data source,}
\par
{\sl Use the Coulomb approximation, of course.}
\par
{\sl The results should certainly be fine}
\par
{\sl For any moderately or highly excited line.}''
\par\noindent
One imagines that the inclusion of this verse made that particular introduction
more widely read than most.

Again a book, rather than a paper, but there's a rather dark introduction in
Goodstein's ``States of Matter'' \cite{States}.
Here are the opening lines of Chapter~1:
``Ludwig Boltzmann, who spent much of his life studying statistical mechanics,
died in 1906, by his own hand.  Paul Ehrenfest, carrying on the work, died
suddenly in 1933.  Now it is our turn to study statistical mechanics.''

For some topics the introductions of papers follow a fairly standard set of
phrases.  As an example, for studies of clusters of galaxies there is usually
mention of how they are the largest virialised structures that exist.  In one
paper I wrote ``All papers on clusters start with a statement about how they
are the largest virialised structures in the Universe, and this paper is
no exception.''  However, my co-authors vetoed this.

\subsection{Contents}
In the main body of his paper,
Carlo Rovelli gave a discussion about the merits of loop quantum gravity
versus string theory in the form of a Socratic dialogue \cite{Rovelli}.
Regarding the overall contents of a paper,
I can't help mentioning ``Chicken Chicken Chicken: Chicken Chicken''
by Doug Zongker \cite{chicken} -- although it's not physics, it might as
well be.   In a paper on ``Relative thermalization'',
the authors wrote ``In order
to keep the above expression only moderately foul \dots We shall spare you the
details (but if you insist, we used \dots and sacrificed a black
chicken)'' \cite{Sacrificed}.  Perhaps unsurprisingly the chicken part didn't
make it into the published version (although the ``foul'' remained).

Although journals tend to insist on formal language, sometimes more
frivolous-sounding statements sneak through the process \cite{Keel}.
A paper on galaxies observed
with the {\it IRAS\/} satellite enumerated the main results in a summary,
concluding with this point: ``{\it IRAS\/} galaxies are all chocolate chip
flavored rather than vanilla flavored as heretofore supposed. This no doubt
accounts for their diversity and appeal'' \cite{Vader}.  In 1981 Fisher \&
Tully stated in the middle of their paper that ``Readers with weak stomachs
may wish to pass to the next subsection'' \cite{Fisher}.

In a paper on quantum entanglement in 2016 \cite{Mahler},
Mahler et al.\  wrote ``The particles
in this article are photons, as was the case in Kocsis et al.''\ then decided
to extend this to ``The particles in this article (Although `the particles in
this article' is in this particular article, consider `the particles in an
article' as part of an article. As any articulate party would know, the
particles in `the particles in an article' are `the' and `in,' whereas the
articles in `the particles in an article' are `the' and `an,' but the
particular article in `the particles in an article' is `the.' `p.s.' is all
that is left when you take the `article' out of `particles.') are photons,
as was the case in Kocsis et al.''  Unfortunately the editors removed
this from the later electronic version of the journal \cite{Wayback}.

Another component of a paper are the figures.  There are
obviously joke figures in joke papers, but there are also examples of
plots in serious papers that are deliberately made to look funny
\cite{pareidolia}.

A particular class of Feynmann diagrams are called ``penguin diagrams''
\cite{penguin1}.
The name originated with John Ellis, and first appeared in a paper as a result
of a bet over a game of darts with Melissa Franklin -- if Ellis lost then he
had to get the word ``penguin'' into his next paper \cite{penguin2}.
He achieved this feat only after realising that the diagrams he had been
studying looked a bit like penguins \cite{Shifman}.

There are many instances where the contents of papers have typos that are
unintentionally comedic.  In cosmology it is surprisingly common to misspell
``redshift'' without the second-last letter; this is usually fixed
by the journal's proof-readers (but not always).
A 1990 paper about the ionized interstellar
medium starts with a statement about the density of ``free elections''
\cite{Reynolds}.  And a prize announcement in 1999 for a certain astronomer
who was an AGN expert referred to his work on ``Anti Galactic Nuclei''.

The lengths of papers vary dramatically.
One of the shortest ever physics papers was ``The Ratio of Proton and Electron
Masses'' by Friedrich Lenz in the Physical Review of 1951 \cite{Pi5}.
The entire content (excluding the single reference) reads:
``The most exact value at present for the ratio of proton to electron mass is
$1816.12\pm0.05$.  It may be of interest to note that this number coincides
with $6\pi^5=1836.12$.''  Unfortunately, as the experimental precision
improved, this numerical coincidence quickly ceased to be consistent with
the data.

In 1981, Hatchett, Begelman \& Sarazin ended their paper on accretion disks
\cite{Hatchett} with this summary:
``Old equations describing disk flex would many a reader perplex, but we've
fixed up some errors and banished the terrors: Our equation is {\it linear\/}
(complex). For a number of torque contributions this allows analytic solutions.
With equal facility we've shown the stability resulting from viscous
diffusions.''

\subsection{Acronyms}
There is a great tradition in physics and astronomy of attaching acronyms to
the names of experiments, projects and other commonly used terms.
Forced and unlovely arrangements of letters seem particularly common in
astronomy, so that recalling your favourite examples has become a kind of
sport \cite{DOOFAAS}.  There is some consensus that the
winner of the ``most awkward acronym'' contest is 11HUGS, which is the
``11 Mpc Halpha and Ultraviolet Galaxy Survey'' \cite{11HUGS}.

Particle physicists also like to make up acronyms for experiments and for
theoretical methods.  GADZOOKS! is the ``Gadolinium Antineutrino Detector
Zealously Outperforming Old Kamiokande, Super!'' (including the exclamation
mark) \cite{GADZOOKS}.
In their ``Chiral Trace Anomalies'' paper of 1973 \cite{TraceAnomalies}
Chanowitz \& Ellis used the abbreviation ``POT'' for ``partially zero trace'',
but the journal objected and suggested ``PZT'', with the compromise solution
``P0T'' appearing in the published version.

Physicists studying dark matter talk about weakly-interacting massive particles
(WIMPs) \cite{WIMPs} and massive compact-halo objects (MACHOs) \cite{MACHOs}.
The more jokey versions are to say that WIMP stands for ``well it might be
particles'' and MACHO is ``maybe astrophysics can help out''.

In spectroscopy we have FASTCARS for ``femtosecond adaptive spectroscopic
techniques for coherent anti-Stokes Raman spectroscopy'' \cite{FASTCARS}.
There is also ``frequency-resolved optical gratings'' (FROG), as well as the
more contrived
French version ``grating-eliminated no-nonsense observation of ultrafast
incident laser light e-fields'' (GRENOUILLE) \cite{FROG}.
The field of nuclear magnetic resonance has many light-hearted acronyms,
such as CAMELSPIN, FLOPSY, HORROR and INEPT \cite{NMR}.

A ``deficient acronym'' might be one where medial letters are sometimes used to
contrive the acronym, rather than just the letters at the starts of the words.
Examples include: ANCHORS, ``AN Archive of CHandra Observations of Regions
of Star formation'' \cite{ANCHORS}; 
FIREFLY, ``Fitting IteRativEly For Likelihood analYsis'' \cite{FIREFLY}; MISS
MARPLE, ``Method for Including Starspots and Systematics in the MARginalized
Probability of a Lone Eclipse'' \cite{MISSMARPLE}; PINOCCHIO,
``PINpointing Orbit-Crossing Collapsed HIerarchial Objects'' \cite{PINOCCHIO};
and SPIDERS, ``SPectroscopic IDentification
of ERosita Sources'' \cite{SPIDERS}.

There are multiple examples of nested acronyms, where part of the acronym is
an acronym itself -- this seems like ``fun with acronyms!''  The ATLAS
experiment is an example, standing for ``A Toroidal LHC ApparatuS''
(which is also ``deficient'', as defined above).  JIVE is the Joint Institute
for VLBI in Europe, while JADES is the JWST Advanced Deep Extragalactic
Survey.

\subsection{Jargon}
Physicists and astronomers are keen on using physics-ese and astronomy-ese
in their papers.  Sometimes the choices of new pieces of technical language
involve a touch of humour.  The names of fundamental particles provide
examples, e.g.\ the neutrino (coined by Amaldi as a joke with Fermi)
and the quark (coined by Gell-Mann, with some influence from James Joyce
\cite{quark}).  And then there are the
supersymmetric particles listed in the Sparticle Data Book, e.g.\
the stop and the wino, plus hypothetical particles, such
as the glueball, the strangelet and the WIMPzilla \cite{WIMPzilla}.

Astronomers were obviously exercising a particular kind of cruel humour when
inventing the ``magnitude'' unit, deciding to call everything heavier than
helium a ``metal'', coining the term ``planetary nebula''
and talking about both ``H{\sc ii} clouds'' and ``H$_2$
clouds''.  Additionally there are jargon words that cause titters among
non-specialists, e.g.\ the adjectives ``degenerate'', ``eccentric'',
``inferior'', ``late'', ``mean'' and ``peculiar''.

The term ``quasar'' was first used in 1964 by Hong-Yee Chiu as an
abbreviated form of ``quasi-stellar radio source'' \cite{quasar}.
The same ``-ar'' suffix was adopted for ``pulsar'' by Bell and Hewish in 1968
and later extended by other researchers to ``blazar'', ``magnetar'',
``collapsar'' and ``blitzar''.  Additional suggestions include
``almucantar'', as well as ``alc{\'a}zar'', ``balthazar'', ``bazaar'',
``hussar'', ``mizar'', ``guitar'' and ``ahoythar'' \cite{pirates}.

``Boojums'', patterns seen in superfluidity, were named after a
nonsense word from Lewis Carroll.  Fluid mechanics contains
delightful invented words like ``enstrophy'' and ``vortensity''.
There are also inadvertently amusing phrases in other branches of physics,
such as Burgers' equation, Killing vectors and Love waves.

\subsection{Units}
Some bizarre units are used in the physical sciences, which could only
have come about through a sense of humour.  Nuclear and
particle physicists use the ``barn'' ($10^{-28}\,{\rm m}^2$) for areas,
from the phrase ``couldn't hit the side of a barn'', and the ``shake''
($10\,$ns) for times, from ``two shakes of a lamb's tail''.
The ``Dirac'' is jokingly defined as a speaking rate of one word per hour
\cite{Dirac}, while
the ``smoot'' is a quirky unit of length, invented as part of a student
prank and named after Oliver Smoot, who fittingly later worked with
organisations that developed standards.  As a fairly unusual surname, it
may not be surprising to learn that Oliver Smoot is a cousin of physicist
George Smoot, who, along with John Mather, won the Nobel Prize in 2006 for
work on the cosmic microwave background \cite{Smoot}.  This was celebrated
in a double-dactyl by mathematician Robin Pemantle:
\par
{\sl Higgeldy Piggeldy}
\par\quad
{\sl Berkeley cosmologist,}
\par\quad
{\sl also a unit of}
\par\quad
{\sl measurement, Smoot,}
\par
{\sl found microscopical}
\par\quad
{\sl anisotropical}
\par\quad
{\sl noises which caused him his}
\par\quad
{\sl own horn to toot.}
\vspace{0.1cm}

When discussing distances on the scale of planetary systems, astronomers use
a length with the imaginative name of the ``astronomical unit''.
Some high-energy astrophysicists use ``foe'' to represent
$10^{51}\,{\rm erg}$ \cite{FOE}, also
sometimes referred to as a ``bethe''.  While the ``hertz'' is the standard
unit for frequency ($\nu$), there is no accepted standard for angular
frequency ($\omega=2\pi\nu$), but it has been suggested that the ``avis''
would be appropriate \cite{avis}.

\subsection{Footnotes}
A 1975 paper by Zuckerman et al.\ on ``Detection of interstellar trans-ethyl
alcohol'' has a ``{\it Note added on proof\/}'' that describes an estimate of
the proof (in terms of alcohol content compared with water) for molecular
clouds \cite{Zuckerman}.

There are also surely many amusing footnotes that have sneaked into papers
and passed into the published versions.  There are probably so many that
it isn't really practical to list examples.\footnote{And I can't think of
any right now.}

\subsection{Acknowledgements}
Numerous examples of jokes are buried in the acknowledgements of
papers.  Often these are sufficiently obscure to be understood only by the
authors or their close colleagues.  For example, there are instances of
grateful thanks given to coffee shops or breweries disguised as the names
of fellow scientists; one example is ``T.\ Cobbold'', for ``Tolly Cobbold'',
a former brewery in England.  The paper by Chodos \& Rabin thanks their
``assistant Beaker, for technical aid and wish him a speedy recovery''
\cite{Chodos}.  The Sokal paper in ``Social Text'' \cite{Sokal} thanks four
individuals ``for enjoyable discussions which have contributed greatly to this
article''; they turn out to be relatives and children of friends, ranging in
age between 2 and 6.  There are also rumours of hidden marriage
proposals, and at least one example of the blunt
phrase ``Will you marry me?''\@ at the end of an acknowledgement \cite{Long}.

A 1976 paper by Chastel, critiquing a speculative
proposal for non-cosmological redshifts, says that the conclusions are
being left to the reader and acknowledges that ``I wrote this paper for
money'' \cite{Chastel}.  In 2004 the lead author of the Sana et al.\ paper
thanks ``the University of Li{\`e}ge for taking care of his integration and for
{\it generously\/} providing heat and electricity'' \cite{Sana}.
In their 2013 paper ``Collective Motion of Humans in Mosh and Circle Pits at
Heavy Metal Concerts'' \cite{Mosh}, the authors make clear that their
``fieldwork was independently funded''.
Authors occasionally also feel the need to
{\it un\/}acknowledge individuals \cite{Goupil}.

\subsection{References}
Spoof papers often contain bogus references.  In the ``Candlestickmaker'' paper
\cite{Candle} all the references are fake, but the journals themselves are real,
e.g.\ ``Trans.\ N.-E.\ Cst.\ Inst.\ Engrs.\ Shipb.'' (the Transactions of the
North-East Coast Institution of Engineers and Shipbuilders) or ``Zentralbl.\
Bakt.'' (Zentralblatt f{\"u}r Bakteriologie).
In 1973 Chanowitz \& Ellis \cite{TraceAnomalies} included a note about
``Dylan's version of Weinberg's theorem'', citing a paper by Zimmerman.

When you submit a paper, it's common to receive complaints from others
for not including particular citations in your
reference list.  In fact the convention involves three distinct steps and
so this is the form letter that I've prepared for these eventualities:
\par\hangindent=10truept\hangafter=1
{\sl Dear Dr. [name here]}
\par\hangindent=10truept\hangafter=1
{\sl Thank you for your interest in our paper on [title], and for
noting that we omitted to cite your own work on a related topic.}
\par\hangindent=10truept\hangafter=1
{\sl However, we should point out that there is a well accepted convention
that is normally followed here, involving three statements that should
be made by the complainant in all such cases:}
\par\hangindent=10truept\hangafter=1
{\sl (1) I enjoyed your paper;}
\par\hangindent=10truept\hangafter=1
{\sl (2) I noticed a minor error in one of the equations;}
\par\hangindent=10truept\hangafter=1
{\sl (3) by the way, you didn't cite me.}
\par\hangindent=10truept\hangafter=1
{\sl Since you omitted the first two of these steps, thus violating the 
established convention, then we will be ignoring your request.}
\par\hangindent=10truept\hangafter=1
{\sl Sincerely,}
\par\hangindent=10truept\hangafter=1
{\sl \quad Douglas Scott [on behalf of the co-authors]}

\subsection{Refereeing}
A last step before a paper is published is the tricky business of refereeing.
``A note on the game of refereeing'' was written by statisticians in 1968
\cite{Game}, but applies equally well to physics.  The basic point of the game
is that authors get more points for publishing pointless papers, while
referees get more points for blocking the publication of worthwhile papers.
Several specific tactics are given, the most effective one for the referee
being to simply ignore all correspondence and delay responding as long as
possible.  When Virginia Trimble was an editor at the Astrophysical Journal,
she would tell people that astronomers were separated into two categories
depending on whether they were fast or slow at refereeing, and that authors
would have their papers sent only to referees in the same category!

An example of the refereeing process in physics comes from a paper submitted
by Krauss in 1986.
The article itself was a spoof of attempts to re-evaluate old data in order to
investigate Newtonian gravity.  It was rejected by Physical Review
Letters, but the most amusing part was that the editors (George Basbas and
perhaps others) decided to respond in kind with six fake referee reports
\cite{FifthFarce}.

\subsection{Postmortems}
After publication comes assessment by the scientific community.  The
literature is full of strongly worded refutations and attacks -- but are
there any genuinely humorously worded rebuttals out there?  Let me pick one
example, which came
in the form of criticism of some of the claims of Immanuel Velikovsky, who
in the 1950s to 1970s wrote pseudohistory and had sensational theories about
catastrophic encounters between the Earth and other planets.  One of the
world's leading experts on the ancient cuneiform script, Abraham Sachs,
said \cite{Sachs}:
``I have read carefully Dr.\ Velikovsky's `Worlds in Collision'
\dots especially carefully those sections -- often quite lengthy -- which deal
with evidence from cuneiform texts, and I have checked all the sources
mentioned in the footnotes. I am happy to report that the bibliographical
references in the footnotes are cited with an amazingly high accuracy. But
having said this, I regret to have to add that I have reported everything that
I can honestly find on the credit side of the ledger. On the negative side,
in the time available, I cannot even list all the errors, misunderstandings,
and false conclusions''.

For books there is the extra step of published reviews, which can certainly be
very harsh at times.  A more light-hearted case appeared in
Nature, in the form of a review (written by Orlando Belpaese)
of the book ``The Bohr-Einstein Transcripts'' by T.J. Gsch{\"a}ftlhuber.
The review describes an early answerphone technology that had been gifted
to Einstein and how newly discovered recordings from the device
included heated conversations between him and Bohr, among more personal
snippets.  This appeared in the issue of 1st April 1993 \cite{Belpaese}.

\section{But seriously}
This review has contained many examples of frivolous contributions to
physics and astronomy.  But is there a point to all of this?
The Monty Python comedy troupe liked to switch topics by using the phrase
``now for something completely different''.  It might seem natural to make such
a statement in order to shift from talking about science to talking about
humour.  However, I'd like to try to convince you that these topics
{\it aren't\/} as different as they might appear; moreover, by
discussing the relationships between them, we might come to see that these
whimsical science contributions actually have some real value.

Pointing out connections between science and humour isn't new \cite{new}.
We've already
mentioned that R.V. Jones wrote about the parallels between the two domains.
In his book ``The Act of Creation'' \cite{Koestler},
novelist and philosopher Arthur Koestler drew a parallel between science
and humour, both involving seeing unexpected connections, which he called
``bisociation'', incorporating the merging of two frames of reference
\cite{Parapsychology}.

Robert P. Crease, philosopher and historian of science (writing in ``Physics
World'' \cite{Crease}), said
``But in a field that uses imagination and play to disclose new truths about
nature \dots the ability to practice both physics and humour are thus
intimately connected -- `entangled', you might say -- inseparably bound up
together in a common and deep-lying origin \dots
only misguided simple pictures of science as a purely logical process relegate
humour to the exterior of the scientific enterprise.''

In 1969 French academic and journalist Robert Escarpit \cite{impact}
expressed the view that a good scientist must have a sense of humour in order
to question beliefs and entertain new concepts and alternative explanations:
``only a sense of humour, then, can guarantee that he remains intellectually
open''.

James McConnell, founder of the Worm Runner's Digest, in the article
``Confessions of a scientific humorist'' \cite{impact} wrote that ``Humour has
no place in Science (capital `S')''; he attempted to define humour, saying
that ``much of it seems a sudden or unexpected departure from the norm, and
that if you don't know what the norm is, the humour is usually lost on you'',
so that specialized science wit requires that the reader brings a lot of
background knowledge.   He ends with the declaration: ``It is my strong
belief that if we can get the younger generation to the point of being able
to laugh at itself, then and only then can we hope to turn
Science back into science.''

The humour described herein consists mostly of in-jokes, which can only be
fully appreciated by people with years of education in the physical
sciences \cite{higher}.  The nature of humour has been debated since the time
of the ancient philosophers and there have been many attempts to explain it.
For example, ethologist Konrad Lorenz \cite{Konrad} said that
laughter is a nervous release from a state of tension.
At the crudest level, some things are funny (like slapstick, for example)
through a feeling of relief (and superiority) over the misfortune of someone
else.  But science paper parodies are {\it not\/} like this -- the humour
doesn't come from enjoying the suffering of a particular other person,
but from feeling superior to {\it everyone\/} outside the group who
understands the joke!  On the other hand,
these in-jokes can serve a positive role in building collegiality.
The old songs from the Cavendish Laboratory are good examples --
they were complimentary about the senior scientists and celebrated physics,
thereby engendering a sense of community among the students.  At the loftiest
end of the spectrum, this then is the goal of science parody.

As discussed at the beginning of this review, the deepest connection between
physics and humour is that both involve
congruities (analogies) and incongruities (discordances).
Isaac Asimov liked to stress that the most important phrase in science
is not ``eureka!''\@ but ``that's funny!''\@ \cite{eureka}, i.e.\ it's the
things that don't quite fit or fit in surprising ways
that lead to forward leaps.
Stumbling across congruity is sometimes what makes the biggest breakthroughs in
physics -- the moments of greatest epiphany are often where one suddenly
realises that some phenomenon is understandable through ideas that at first
seem completely unrelated.  Examples include: seeing the same equations for
AC circuits as for springs; the unification of electricity and magnetism
yielding photons; interpreting gravity through pure geometry; the connection
between entropy and information; the thermodynamics of black holes;
seeing critical phenomena in quite different physical systems;
and more recently the AdS/CFT correspondence.  Each reader probably has their
own favourite examples.  The point is that these moments of connection have
a lot in common with the realisation that something is funny.

So how does one find these breakthroughs in physical understanding?  They are
surely enabled by thinking ``outside the box'', imagining different
kinds of explanation, including those that might at first seem ridiculous.
I would claim that a similar thought process goes on when
physicists make important new connections as is happening in the minds of
great comedians.

Let me add one other further thought: the world could use more humour!  To
employ a cosmological analogy, the Universe is dominated by a mysterious
substance usually referred to as ``dark energy'', but as pointed out many
times, this is a bit of a misnomer.  The name doesn't emphasize the bizarre
equation of state, which involves a negative pressure, leading to acceleration
in the scale factor of the Universe.  The name is also obviously a bit
``dark'', emphasising grimness, obscurity and gloom!  An
alternative suggestion is to call it ``levity'' \cite{levity}.  Apart from
being a more appropriate name, I like the
idea that the most important constituent of the Cosmos is levity.

\section{Conclusions}
There are no conclusions
\cite{further}\cite{thanks}\cite{fake}\cite{apologies}\cite{conclusions}.


\smallskip


\baselineskip=11pt

\section*{Appendix~A. Ig Nobel Prizes}
Ig Nobel Prizes are given regularly (although not every year) for Physics,
with the winners including these papers:
\vspace{0.1cm}
\begin{itemize}
\setlength\itemsep{3pt}
\item ``The heaviest element in the universe, Administratium'', 1994
\cite{Ig1991};
\item ``A Study of the Effects of Water Content on the Compaction Behaviour
of Breakfast Cereal Flakes'', 1995 \cite{Ig1995};
\item ``Tumbling toast, Murphy's Law and the fundamental constants'', 1996
\cite{Ig1996};
\item ``Physics Takes the Biscuit'', 1999 \cite{Ig1999};
\item ``Of Flying Frogs and Levitrons'', 2000 \cite{Ig2000};
\item ``Demonstration of the Exponential Decay Law Using Beer Froth'', 2002
 \cite{Ig2002};
\item ``An Analysis of the Forces Required to Drag Sheep over Various
Surfaces'', 2003  \cite{Ig2003};
\item ``Coordination Modes in the Multisegmental Dynamics of Hula Hooping'',
2004  \cite{Ig2004};
\item ``The Pitch Drop Experiment'', 2005  \cite{Ig2005};
\item ``Fragmentation of Rods by Cascading Cracks: Why Spaghetti Does Not
Break in Half'', 2006 \cite{Ig2006};
\item ``Geometry and Physics of Wrinkling'', 2007 \cite{Ig2007};
\item ``Spontaneous Knotting of an Agitated String'', 2008 \cite{Ig2008};
\item ``Shape of a Ponytail and the Statistical Physics of Hair Fiber
Bundles'', 2012 \cite{Ig2012};
\item ``Humans Running in Place on Water at Simulated Reduced Gravity'', 2013
\cite{Ig2013};
\item ``Frictional Coefficient under Banana Skin'', 2014 \cite{Ig2014};
\item ``On the Rheology of Cats'', 2017 \cite{Ig2017};
\item ``How Do Wombats Make Cubed Poo?'', 2019 \cite{Ig2019};
\item ``Excitation of Faraday-like body waves in vibrated living earthworms'',
2020 \cite{Ig2020}.
\end{itemize}

\vspace{0.1cm}
Some earlier prizes were given derisively to pieces of pseudo-science, but the
more recent awards are for genuine scientific studies that might just
{\it seem\/} to be ridiculous, but in fact demonstrate something interesting.
Disregarding those awarded for fringe science claims (like the face on
Mars and ancient astronauts), there has been only one prize related to
astronomy:
\vspace{0.1cm}
\begin{itemize}
\setlength\itemsep{3pt}
\item ``Dung Beetles Use the Milky Way for Orientation'', 2013 \cite{IgAstro}.
\end{itemize}

\section*{Appendix~B. April Fool's APODs}
Here's a list of Astronomical Pictures of the Day with an April Fool's Day
theme \cite{APOD}:
\vspace{0.1cm}
\begin{itemize}
\setlength\itemsep{3pt}
\item
``Ski Mars!'', 1999;
\item
``A New Constellation Takes Hold'', 2003;
\item
``April Fools Day More Intense On Mars'', 2004;
\item
``Water On Mars'', 2005;
\item
``Hubble Resolves Expiration Date For Green Cheese Moon'', 2006;
\item
``Americans Defeat Russians in First Space Quidditch Match'', 2007;
\item
``New Space Station Robot Asks to be Called `Dextre the
Magnificent'\thinspace'', 2008;
\item
``Astronaut's Head Upgraded During Spacewalk'', 2009;
\item
``Evidence Mounts for Water on the Moon'', 2010;
\item
``It's Raining on Titan'', 2011;
\item
``Dad Quiets Omicron Ceti'', 2012;
\item
``Moon or Frying Pan?'', 2013;
\item
``Space Station Robot Forgets Key Again'', 2014;
\item
``Suiting Up for the Moon'', 2015;
\item
``Europa: Discover Life Under the Ice'', 2016;
\item
``Split the Universe'', 2017;
\item
``I Brought You the Moon'', 2018;
\item
``Astronaut Kicks Lunar Field Goal'', 2019;
\item
``Asteroid or potato?'', 2020.
\end{itemize}

\section*{Appendix~C.  ArXiv April Fools}
This is a list of ``April Fool'' type papers submitted to the eprint arXiv.
Although lengthy, the list is undoubtedly incomplete.  In chronological
order they are:
\vspace{0.1cm}
\begin{itemize}
\setlength\itemsep{3pt}
\item
``Superiority of the Lunar and Planetary Laboratory (LPL) over Steward
Observatory (SO) at the University of Arizona'', 2002 \cite{arXiv0204013};
\item
``On the Utter Irrelevance of LPL Graduate Students: An Unbiased Survey by
Steward Observatory Graduate Students'', 2002 \cite{arXiv0204041};
\item
``Cosmic Conspiracies'', 2006 \cite{Frolop1};
\item
``The Stryngbohtyk Model of the Universe: a Solution to the Problem of the
Cosmological Constant'', 2007 \cite{arXiv0703774};
\item
``Natural Dark Energy'', 2007 \cite{Frolop2};
\item
``On the origin of the cosmic microwave background anisotropies'', 2007
\cite{arXiv0703806};
\item
``Relativity Revisited'', 2008 \cite{arXiv08040016};
\item
``Down-sizing Forever'', 2008 \cite{Frolop3};
\item
``Time variation of a fundamental dimensionless constant'', 2009
\cite{arXiv09035321};
\item
``Galaxy Zoo: an unusual new class of galaxy cluster'', 2009
\cite{arXiv09035377};
\item
``Orthographic Correlations in Astrophysics'', 2010 \cite{arXiv10036064};
\item
``Schroedinger's Cat is not Alone'', 2010 \cite{arXiv10044206};
\item
``The Observed Inclination Problem: Solved at Last?'', 2011
\cite{arXiv11036167};
\item
``Non-standard morphological relic patterns in the cosmic microwave
background'', 2011 \cite{ZZZZ};
\item
``On the influence of the Illuminati in astronomical adaptive optics'', 2012
\cite{arXiv12036708};
\item
``Gods as Topological Invariants'', 2012 \cite{arXiv12036902};
\item
``The Proof of Innocence'', 2012 \cite{arXiv12040162};
\item
``On the Ratio of Circumference to Diameter for the Largest Observable
Circles: An Empirical Approach'', 2012 \cite{arXiv12040298};
\item
``Non-detection of the Tooth Fairy at Optical Wavelengths'', 2012
\cite{arXiv12040492};
\item
``Pareidolic Dark Matter (PaDaM)'', 2013 \cite{arXiv13037262};
\item
``A search for direct heffalon production using the ATLAS and CMS experiments
at the Large Hadron Collider'', 2013 \cite{arXiv13037367};
\item
``Felinic principle and measurement of the Hubble parameter'', 2013
\cite{arXiv13037382};
\item
``Unidentified Moving Objects in Next Generation Time Domain Surveys'', 2013
\cite{arXiv13037433};
\item
``Conspiratorial cosmology - the case against the Universe'', 2013
\cite{arXiv13037476};
\item
``Empirical Limits on the Russell Conjecture'', 2013 \cite{arXiv13040240};
\item
``Winter is coming'', 2013 \cite{arXiv13040445};
\item
``The CMB flexes its BICEPs while walking the Planck'', 2014 \cite{Frolop4};
\item
``Bayesian Prediction for The Winds of Winter'', 2014 \cite{arXiv14095830};
\item
``A Farewell to Falsifiability'', 2015 \cite{Frolop5};
\item
``Beyond the New Horizon: The Future of Pluto'', 2015 \cite{arXiv150400630};
\item
``SET-E: The Search for Extraterrestrial Environmentalism'', 2016
\cite{arXiv160309428};
\item
``Astrology in the Era of Exoplanets'', 2016 \cite{arXiv160309496};
\item
``An unexpected new explanation of seasonality in suicide attempts: Grey's
Anatomy broadcasting'', 2016 \cite{arXiv160309590};
\item
``Pi in the sky'', 2016 \cite{Frolop6};
\item
``Pipe-cleaner Model of Neuronal Network Dynamics'', 2016 \cite{arXiv160309723};
\item
``Stopping GAN Violence: Generative Unadversarial Networks'', 2017
\cite{arXiv170302528};
\item
``Detecting the Ultimate Power in the Universe with LSST'', 2017
\cite{arXiv170310432};
\item
``A Neural Networks Approach to Predicting How Things Might Have Turned Out
Had I Mustered the Nerve to Ask Barry Cottonfield to the Junior Prom Back in
1997'', 2017 \cite{arXiv170310449};
\item
``On the Impossibility of Supersized Machines'', 2017 \cite{arXiv170310987};
\item
``Independent Discovery of a Sub-Earth in the Habitable Zone Around a Very
Close Solar-Mass Star'', 2018 \cite{arXiv180400419};
\item
``Super-Earths in need for Extremly Big Rockets'', 2018
\cite{arXiv180311384};
\item
``Sitnikov in Westeros: How Celestial Mechanics finally explains why winter is
coming in Game of Thrones'', 2018 \cite{arXiv180311390};
\item
``Colonel Mustard in the Aviary with the Candlestick: a limit cycle attractor
transitions to a stable focus via supercritical Andronov-Hopf bifurcation'',
2018 \cite{arXiv180311559};
\item
``ACRONYM: Acronym CReatiON for You and Me'', 2019 \cite{arXiv190312180};
\item
``Fast Radio Bursts from Terraformation'', 2019 \cite{arXiv190312186};
\item
``The Long Night: Modeling the Climate of Westeros'', 2019
\cite{arXiv190312195};
\item
``Superfluous Physics'', 2019, \cite{superfluous};
\item
``A new kind of radio transient: ERBs'', 2019 \cite{Frolop7};
\item
``Worlds in Migration'', 2019 \cite{arXiv190312437};
\item
``Forecasting Future Murders of Mr. Boddy by Numerical Weather Prediction'',
2019 \cite{arXiv190312604};
\item
``The Marshland Conjecture'', 2019 \cite{Marsh};
\item
``Cosmological Dark Matter: a Review (the April Fool Edition)'', 2020
\cite{arXiv200313696};
\item
``Quantum Godwin's Law'', 2020 \cite{arXiv200313715};
\item
``Defining the Really Habitable Zone'', 2020 \cite{arXiv200313722};
\item
``Making It Rain: How Giving Me Telescope Time Can Reduce Drought'', 2020
\cite{arXiv200313879};
\item
``Resolving Exo-Continents with Einstein Ring Deconvolution'', 2020
\cite{arXiv200313918};
\item
``The search for life and a new logic'', 2020 \cite{Frolop8};
\item
``An Artificially-intelligent Means to Escape Discreetly from the Departmental
Holiday Party; guide for the socially awkward'', 2020 \cite{arXiv200314169};
\item
``Novel approach to Room Temperature Superconductivity problem'', 2020
\cite{arXiv200314321};
\item
``A PDF PSA, or Never gonna set\_xscale again -- guilty feats with
logarithms'', 2020 \cite{arXiv200314327};
\item
``Searching for Space Vampires with TEvSS'', 2020 \cite{arXiv200314345};
\item
``Conspiratorial cosmology. II. The anthropogenic principle'', 2020
\cite{arXiv200400401};
\item
``Using Artificial Intelligence to Shed Light on the Star of Biscuits:
The Jaffa Cake'', 2021 \cite{arXiv210316575};
\item
``Detection of Rotational Variability in Floofy Objects at Optical
Wavelengths'', 2021 \cite{arXiv210316636};
\item
``The secret of the elixir of youth of blue straggler stars'', 2021
\cite{arXiv210316866};
\item
``Pandemic Dark Matter'', 2021 \cite{arXiv210316572};
\item
``\thinspace`I'll Finish It This Week' And Other Lies'', 2021;
\cite{arXiv210316574}
\item
``The Swampland Conjecture Bound Conjecture'', 2021 \cite{arXiv210316583};
\item
``I Knew You Were Trouble: Emotional Trends in the Repertoire of Taylor
Swift'', 2021 \cite{arXiv210316737};
\item
``My cat Chester's dynamical systems analysyyyyy7777777777777777y7is of the
laser pointer and the red dot on the wall: correlation, causation, or
SARS-Cov-2 hallucination?'', 2021 \cite{arXiv210317058};
\item
``The Existential Threat of Future Exoplanet Discoveries'', 2021
\cite{arXiv210317079};
\item
``The Swapland'', 2021 \cite{arXiv210317198}.
\end{itemize}

\section*{Appendix~D. Funny paper titles}
Here are a few examples of jokey paper titles, selected in an entirely
subjective way, in roughly chronological order \cite{incomplete}:
\vspace{0.1cm}
\begin{itemize}
\setlength\itemsep{3pt}
\item
``Deuteronomy. Synthesis of Deuterons and the
Light Nuclei during the Early History of the Solar System'' by
Fowler, Greenstein \& Hoyle \cite{Deut};
\item
``My World Line: An Informal Autobiography'' by Gamow \cite{WorldLine};
\item
``Can one tell QCD from a hole in the ground?'' by De R{\'u}jula, Ellis,
Petronzio, Preparata \& Scott \cite{Hole1};
\item
``Superspace aspects of supersymmetry and supergravity'' by Ferrara
\cite{Ferrara};
\item
``Axions: To be or not to be?'' by Barroso \& Mukhopadhyay \cite{ToBe};
\item
``Constitutive laws, tensorial invariance and chocolate cake'' by Rundle
\& Passman \cite{Constitutive};
\item
``The effect of birds on radio astronomy'' by Partridge, Peacock \& Gull
\cite{Birds};
\item
``Cosmic Voids: Much Ado About Nothing'' by Gregory \cite{Ado};
\item
``The sphaleron strikes back: A response to objections to the sphaleron
approximation'' by Arnold \& McLerran \cite{Strikes};
\item
``What do you get if you multiply six by nine'' by Adams \cite{Adams};
\item
``A Case for $H_0=42$ and $\Omega_0=1$ Using Luminous Spiral Galaxies and the
Cosmological Time Scale Test'' by Sandage \cite{Sandage};
\item
``Escape from the Menace of the Giant Wormholes'' by Coleman \& Lee
\cite{ColemanLee};
\item
``Effective Lagrangians for p-branes'' by Amorim \& Barcelos-Neto
\cite{PBranes};
\item
``CCD Data: The Good, The Bad, and the Ugly'' by Massey \& Jacoby
\cite{CCD};
\item
``Galaxies form at peaks -- Not!'' by Katz, Quinn \& Gelb \cite{Not};
\item
``Is a local bar a good place to find a companion? The near infrared
morphology of Maffei 2'' by Hurt, Merrill, Gatley \& Turner \cite{LocalBar};
\item
``Supernatural inflation'' by Randall, Soljacic \& Guth \cite{Supernatural};
\item
``$H_0$: The Incredible Shrinking Constant, 1925--1975'' by Trimble
\cite{Shrinking};
\item
``Is string theory a theory of strings?'' by Johnson et al.\ \cite{IsString};
\item
``Cosmic Strings -- Dead Again?'' by Hindmarsh \cite{DeadAgain};
\item
``Anatomy of a Duality'' by Johnson \cite{Anatomy};
\item
``10=6+4'' by Smith \cite{arXiv9908205};
\item
``Raiders of the Lost AdS'' by Kumar \cite{Raiders};
\item
``Why the universe is just so'' by Hogan \cite{JustSo};
\item
``Warped Phenomenology'' by Davoudiasl, Hewett \& Rizzo \cite{Warped};
\item
``Cloudshine: New Light on Dark Clouds'' by Foster \& Goodman \cite{Sunshine};
\item
``Brane New World'' by Hawking, Hertog \& Reall \cite{Brane};
\item
``Boomerang returns unexpectedly'' by White, Pierpaoli \& Scott
\cite{boomerang};
\item
``Domain walls in supersymmetric QCD: The taming of the zoo'' by Binosi \&
Ter Veldhuis \cite{Taming};
\item
``Don't panic! closed string tachyons in ALE spacetimes'' by Adams, Polchinski
\& Silverstein \cite{DontPanic};
\item
``Decapitating the Duck'' by Thorsett, Brisken \& Goss \cite{Duck};
\item
``A Phantom Menace?'' by Caldwell \cite{phantom};
\item
``Living with Ghosts'' by Hawking \& Hertog \cite{Ghosts2};
\item
``Nutty Bubbles'' by Ghezelbash \& Mann \cite{Nutty};
\item
``Brane Big-Bang Brought by Bulk Bubble'' by Gen, Ishibashi \& Tanaka
\cite{BulkBubble};
\item
``One ring to encompass them all: a giant stellar structure that surrounds
the Galaxy'' by Ibata et al.\ \cite{Ibata};
\item
``For whom the disc tolls'' by Lasota \cite{DiscTolls};
\item
``X \& Y'' by Maiani et al.\ \cite{XandY};
\item
``The tachyon at the end of the universe'' by McGreevy \& Silverstein
\cite{Tachyon};
\item
``A Fly in the SOUP'' Holman \& Mersini-Houghton \cite{Fly};
\item
``$\log(M_{\rm Pl}/m_{3/2})$'' by Loaiza-Brito et al.\ \cite{Log};
\item
``Why Eppley and Hannah's Experiment Isn't'' by Mattingly \cite{Isnt};
\item
``And Don't Forget the Black Holes'' by Bethe, Brown \& Lee \cite{DontForget};
\item
``How Much Mass Do Supermassive Black Holes Eat in Their Old 
Age?'' by Hopkins, Narayan \& Hernquist \cite{OldAge};
\item
``It's a gluino!'' by Alves, Eboli \& Plehn \cite{Gluino};
\item
``Does Smoothing Matter?'' by Martin \& van Nieuwenhuizen \cite{Smoothing};
\item
``Elements, topology and T-shirts'' by Fraundorf \cite{Elements};
\item
``Walking in the SU(N)'' by Dietrich \& Sannino \cite{SUN};
\item
``The Matrix Reloaded -- on the Dark Energy Seesaw'' by Enqvist, Hannestad
\& Sloth \cite{Matrix};
\item
``Would Bohr be born if Bohm were born before Born?'' by Nikoli{\'c}
\cite{BohrBohm};
\item
``27/32'' by Tachikawa \& Wecht \cite{Tachikawa};
\item
``Turduckening black holes: An analytical and computational study'' by Brown
et al.\ \cite{Turdecken};
\item
``Velocity dispersions in a cluster of stars: How fast could Usain Bolt have
run?'' by Eriksen, Kristiansen, Langangen \& Wehus \cite{Bolt};
\item
``\thinspace`Kerrr' black hole: The Lord of the string'' by Smailagic \&
Spallucci \cite{Kerrr};
\item
``Simple exercises to flatten your potential'' by Dong, Horn, Silverstein \&
Westphal \cite{Dong};
\item
``Resolving the Radio Source Background: Deeper Understanding through
Confusion'' by Condon et al.\ \cite{Condon};
\item
``Fab Four: When John and George Play Gravitation and Cosmology'' by
Bruneton et al.\ \cite{FabFour};
\item
``The Unbearable Beingness of Light, Dressing and Undressing Photons in Black
Hole Spacetimes'' by Hollowood \& Shore \cite{Unbearable};
\item
``Unconstraining the unHiggs model'' by Englert et al.\ \cite{UnHiggs};
\item
``Close encounters of the protostellar kind in IC 1396N'' by Beltran et al.\
\cite{Close};
\item
``Catastrophic Consequences of Kicking the Chameleon'' by Erickcek, Barnaby,
Burrage \& Huang \cite{Kicking};
\item
``Conservative constraints on early cosmology with MONTE PYTHON'' by
Audren, Lesgourgues, Benabed \& Prunet \cite{MONTE};
\item
``Life, the Universe, and everything -- 42 fundamental questions'' by
Allen \& Lidstr{\"o}m \cite{LUE};
\item
``Some Generalities about Generality'' by Barrow \cite{Generality};
\item
``Hot spaghetti: Viscous gravitational collapse'' by M{\"u}ller \&
Sch{\"a}fer \cite{Spaghetti};
\item
``Snakes on a Spaceship -- An Overview of Python in Heliophysics'' by Burrell
et al.\ \cite{Snakes};
\item
``To $B$ or not to $B$: Primordial magnetic fields from Weyl anomaly'' by
Benevides, Dabholkar \& Kobayashi \cite{ToBB};
\item
``Fisher for complements: extracting cosmology and neutrino mass from the
counts-in-cells PDF'' by Uhlemann et al.\ \cite{Uhlemann};
\item
``Pancakes as opposed to Swiss cheese'' by N{\'a}jera \& Sussman
\cite{PancakeCheese};
\item
``The Hubble Tension Bites the Dust: Sensitivity of the Hubble Constant
Determination to Cepheid Color Calibration'' by Mortsell et al.\ \cite{Tension}.
\end{itemize}


\begin{thebibliography}{9}
\bibitem{tshirt} And the other t-shirt that says ``Astronomy is out of this
world, and looking up, and also amusing''.
\bibitem{Earth} Not that I would suggest for a minute that physicists are
more conceited and arrogant than the general population.
\bibitem{Franklin} When asked what he uttered when he accidentally electrocuted
himself when trying to kill a turkey using a series of Leyden jars, he replied
``I didn't say anything, I was too shocked''.
\bibitem{notket} One of these tends to raise giggles from undergraduates when
they first come across this notation (hint: it's not ``ket'').
\bibitem{nice} It may be worth pointing out explicitly that, like all good
lampoons, this paper mocks an idea rather than an individual.
\bibitem{German} Presumably the ambiguity of language works just as well in
English as it did in the original German.
\bibitem{lions} Admittedly this is probably not the funniest prank of all
time, but it was early days for April Fools tricks.
\bibitem{Poisson} There are variants in other countries also.
In France one shouts ``Poisson d'avril'' after successfully
pranking someone; no one knows the origin of this phrase, but it has been
suggested to be named after the famous French mathematician and physicist
Sim{\'e}on Poisson, who was a noted jokester.
\bibitem{Daftie} Additionally, at school Maxwell was known
as ``Daftie'', essentially because of his strong rural accent.
\bibitem{hoax} Said to be reprinted from the Edinburgh Journal of Science,
which does not in fact exist.  It may be significant that these discussions
of the abundance of life appeared in ``The Sun''.
\bibitem{Lions} In other words, no animals were harmed in the making of this
paper.
\bibitem{threshold} This is sometimes referred to as the worst episode of
Star Trek Voyager, the plot involving an attempt to reach Warp 10, leading to
crew members reverting to a primitive amphibian state.
\bibitem{CatJokes} This is probably why most physicists find essentially none
of the popular jokes involving Schr{\"o}dinger's cat to be funny!
\bibitem{Delter} Gamow also jokingly wrote that their other collaborator,
Robert Herman, ``stubbornly refuses to change his name to Delter''.
\bibitem{Charfman} It does sound vaguely rude though.  Other fake authors have
been invented
that sound even ruder in either English or a different language; we leave these
for curious readers to investigate for themselves.
\bibitem{elk} This might be called a ``moose'' in North America, where an
``elk'' is a different thing entirely, also known as a ``wapiti''.
The different names for these animals can be confusing.  Fortunately
``Tycho Brahe'' has the same name everywhere.
\bibitem{starwars} Although the papers about the proposed missile
defence system tend to be rather less funny.
\bibitem{pareidolia} As well as images that look like other things --
pareidolia.
\bibitem{penguin2} The bet appears to have been one-sided, so we don't know
what would have happened if Ellis had won the darts game.  Perhaps we
would now have a ``parrot'' equation or a new particle called the ``pigeon''.
\bibitem{NMR} Leaving out the ones that may not be appropriate for a
family show; however, they may be available on special request.
\bibitem{quark} And not taken seriously in Germany, since there it's cheese
curd, and also means ``rubbish'', a bit like ``tripe'' in Britain.  The names
of the six quark flavours are also hard to take entirely seriously.
\bibitem{Dirac} David Dewhirst, astronomy librarian in Cambridge, liked to
tell the following story.  On attending an annual college dinner, he
found himself sitting next to Dirac.  Struggling for a way to converse with
the great man he asked ``Do you come here every year?'', to which there was
no reply and the meal proceeded in silence.  However, at the end of the dinner
Dirac turned to him and said ``No''.  Dewhirst thereby boasted that he'd had a
conversation with Dirac that lasted the whole evening.
\bibitem{Smoot} A joke made at the time was that Mather and Smoot won the
prize for studying whether ``matter is smooth''.
\bibitem{FOE} Although in SI-derived units this would be $10^{44}\,{\rm J}$,
which gives the more awkward-sounding ``ffj''.
\bibitem{new} Although not relevant, let us recall that ``humours''
were historically the fluids of the body whose study was connected with the
development of medical science; imbalance in these fluids was believed to
cause temperament and eccentricity.
\bibitem{higher} Someone once said that the point of higher education was so
that you could understand more jokes.
\bibitem{Konrad} Third cousin twice removed of Hendrik Ludvig.  Another distant
cousin, Edward Lorenz invented chaos theory; it is interesting to note that
the phrase ``butterfly effect'' came from a conference organiser giving his
talk the joke title ``Predictability: Does the Flap of a Butterfly's Wings in
Brazil Set Off a Tornado in Texas?'' when Lorenz failed to submit one himself.
\bibitem{eureka} Where he meant ``funny peculiar'', not ``funny ha-ha''.
\bibitem{fake} In the spirit of April Fool's Day, there are perhaps a handful
of things in this article that may not be entirely factually correct.  Can you
spot them all?
\bibitem{incomplete} Of course this list is far from complete, and entirely
subjective.
\end{thebibliography}
\end{document}